\documentclass[twocolumn]{openjournal}
\usepackage{graphicx}   
\usepackage{amsmath}    
\usepackage{amssymb,amsfonts}    
\usepackage{orcidlink}
\usepackage{verbatim}
\usepackage{hyperref}
\usepackage{xcolor}

\shorttitle{I. Waisberg}
\shortauthors{I. Waisberg}

\begin{document}

\title{Which is the most eccentric binary known? \\ Insights from the 2023/4 pericenter passages of Zeta Bo\"otis and Eta Ophiuchi}

\footnotetext[]{Based on observations collected at the European Southern Observatory, Chile, Program IDs 111.24UP.001, 111.264P.001 and 113.26HF.001}

\newcommand{\weizmann}{Department of Particle Physics and Astrophysics, Weizmann Institute of Science, Rehovot 76100, Israel}

\email{email: idelwaisberg@gmail.com}

\author{\vspace{-1.2cm}Idel Waisberg\,\orcidlink{0000-0003-0304-743X}$^{1,2}$, Ygal Klein\orcidlink{0009-0004-1914-5821}$^{2}$ \& Boaz Katz\orcidlink{0000-0003-0584-2920}$^{2}$}

\affiliation{$^1$Independent researcher}
\affiliation{$^2$\weizmann}

\begin{abstract}
There is a clear dearth of very eccentric binaries among those for which individual eccentricities can be measured.  In this paper we report on observations of the two nearby, bright and very eccentric visual binaries Zeta Bo\"otis ($\zeta$ Boo) and Eta Ophiuchi ($\eta$ Oph), for which VLTI/GRAVITY interferometric observations were obtained during their pericenter passages in 2023/4.  Previous observations of $\zeta$ Boo suggest an eccentricity $e>0.99$ with high significance, implying that it has the highest eccentricity of any known binary. However, our interferometric measurements near periastron passage reveal that the eccentricity is actually $e=0.980450\pm0.000064$ (second highest well constrained eccentricity) with a pericenter distance $a_p=0.818\pm0.009\text{ au}$. We attribute the previous over-estimation to a degeneracy that plagues very eccentric visual binary orbital solutions. For $\eta$ Oph we find an eccentricity $e=0.93077\pm0.00013$ (compared to previous estimates of $e=0.95\pm 0.02$), a pericenter distance $a_p=2.15\pm0.10 \text{ au}$ and attribute the over-estimated dynamical mass in the previous solution to an underestimated error in the semi-major axis. With no additional close companions capable of influencing their further evolution, both systems are expected to fully circularize as the stars evolve and expand, ultimately leading to close circular binaries with no memory of their very eccentric past. 
\end{abstract}


\section{Introduction}
\label{sec:introduction}

Several direct and indirect lines of evidence suggest that wide stellar binaries (periods $P\gtrsim 1000 \rm d$) appear to have large eccentricities with a distribution that is flat or rising as $e$ approaches unity \citep[e.g.][]{Duquennoy91,Tokovinin20,Hwang22,Offner23}. However, while about $\gtrsim 5\%$ of such binaries are expected to have extreme $e\ge 0.95$ eccentricities, very few such orbits are known with high confidence. Focusing on periods $P>500$ days, for which tidal circularization is negligible for main sequence stars, and on orbital solutions with full coverage and small reported eccentricity errors of $<0.05$, there are 5 such systems out of $\sim 50,000$ in Gaia DR3 \citep{Gaia23,Gaia23b,Gaia23c}, 4 out of $\sim 400$ in the Sixth Catalog of Orbits of Visual Binary Stars\footnote{\href{http://www.astro.gsu.edu/wds/orb6.html}{http://www.astro.gsu.edu/wds/orb6.html}} (hereafter ORB6) and 2 out of $\sim 1,000$  in the 9th Catalogue of Spectroscopic Binary Orbits \citep[SB9; ][]{Pourbaix04}. The dearth of high-eccentricity orbits is often attributed to selection biases \citep[e.g.][]{Harrington77, Tokovinin16}, but the fact remains that we know very little about the actual distribution at high eccentricities $e>0.9$.

Zeta Bo\"otis and Eta Ophiuchi are two nearby, bright and very eccentric $e \gtrsim 0.95$ visual binaries that happened to undergo pericenter passage in 2023/4. Given their relatively long periods of 125 and 88 yrs, this presented a unique opportunity to observe their pericenter passage with long-baseline interferometry in order to better constrain their orbital parameters. The stellar components of both pairs are intermediate mass A-type main sequence stars. 

Zeta Bo\"otis ($\zeta$ Boo, HIP 71795, HD 129246/7, HR 5477/8, 30 Boo) is a bright (V=3.8) visual binary (WDS J14411+1344). The secondary (B) is almost as bright as the primary (A) with $\Delta V_T = 0.04$ \citep{Fabricius00}. The combined spectral classification is A1V in \cite{Abt95} and the projected rotational velocity is $v \sin i = 156 \text{ km}\text{ s}^{-1}$ \citep{Abt2004}. The Hipparcos parallax \citep[$p = 18.07 \pm 1.24 \text{ mas}$; ][]{Hipparcos97} implies a distance $d=55 \pm 4 \text{ pc}$ ($\zeta$ Boo is unsolved in Gaia DR3). 

The current orbital solution for $\zeta$ Boo in ORB6 is that from \cite{Scardia07} with a period $P=125.54 \text{ yr}$, a semi-major axis $a=0.825"$, an eccentricity $e=0.98$ and an inclination $i=126.0^{\degr}$ (no uncertainties are reported). The resulting dynamical mass $M_{\mathrm{dyn}} = 4.93 M_{\odot}$ (using the more precise distance of its wide companion; see paragraph below) is reasonably consistent with that expected for the pair of A-type stars. However, there are two alternative orbital solutions for $\zeta$ Boo listed in the master file of ORB6 with the same quality but higher eccentricities. A solution by \cite{Alzner07} has $P=122.98 \text{ yr}$, $a=0.892"$ ($M_{\mathrm{dyn}} = 6.5 M_{\odot}$), $e=0.985$ and $i=126.0^{\degr}$. More intriguingly, a more recent solution in \cite{Muterspaugh10a}, based on combining their observations of $\zeta$ Boo with the Palomar Testbed Interferometer with the previous measurements, has $P=124.46 \pm 0.17 \text{ yr}$, $a=2.3 \pm 1.7"$ ($M_{\mathrm{dyn}} = 110 \pm 470 M_{\odot}$), $e=0.9977 \pm 0.0034$ and $i=102.3 \pm 9.2^{\degr}$. Such a high eccentricity would make $\zeta$ Boo the most eccentric binary known. One of the goals of this paper is to understand the origin of such discrepancies even for a star as well observed as $\zeta$ Boo, with about 750 astrometric measurements covering two revolutions since 1796. 

In \cite{Waisberg24} we reported on the discovery that the F0V star HIP 71759 (HD 129153, HR 5473) is a very wide companion to $\zeta$ Boo at a projected separation $\rho = 13.3' \leftrightarrow 41.3 \text{ kau} = 0.20 \text{ pc}$. The association was based on their consistent parallaxes and proper motions, with a proper motion difference $\Delta v_{\mathrm{2D}} = 0.86 \pm 0.20 \text{ km} \text{ s}^{-1}$ compared to the 2D Keplerian escape velocity $v_{\mathrm{esc,2D}} \approx 0.50 \text{ km} \text{ s}^{-1}$. Given the smaller uncertainty, we adopt the Gaia DR3 distance to the very wide companion HIP 71759 \citep[$p =19.33\pm0.05 \text{ mas} \leftrightarrow d=51.7 \pm 0.1 \text{ pc}$; ][]{Gaia23} also for $\zeta$ Boo. We increase the distance error of $\zeta$ Boo to 0.2 pc to account for the large projected separation between them. 

Eta Ophiuchi ($\eta$ Oph, HIP 84012, HD 155125, HR 6378, 35 Oph) is another bright (V=2.4),  nearby very eccentric visual binary (WDS J17104-1544). The primary (A, with the IAU-approved proper name \textit{Sabik}) is brighter by $\Delta V_T = 0.47$ \citep{Fabricius00} than the secondary (B). The combined spectrum was classified as A2IV-V in \cite{Gray06} and has a rather low projected rotational velocity ($v \sin i = 10 \text{ km} \text{ s}^{-1}$) and no evidence for any spectral peculiarity \citep{Catanzaro06}. The Hipparcos parallax $p = 38.77 \pm 0.86 \text{ mas}$ \citep{Hipparcos97} translates to a distance $d=25.8 \pm 0.6 \text{ pc}$, which we adopt throughout this paper ($\eta$ Oph is unsolved in Gaia DR3).

The current orbital solution for $\eta$ Oph listed in ORB6 is that from \cite{Docobo07} with a period $P=87.58 \pm 1.00 \text{ yr}$, a semi-major axis $a=1.396 \pm 0.010"$, an eccentricity $e=0.95 \pm 0.02$ and an inclination $i=95.2 \pm 2.0^{\degr}$. The resulting dynamical mass $M_{\mathrm{dyn}} = 6.09 \pm 0.44 M_{\odot}$ is somewhat too high for a pair of A2 dwarfs and could be explained by either higher multiplicity or underestimated errors in the orbital parameters. 

This paper is organized as follows. In Section \ref{sec:results}, we report on our interferometric observations of the two binaries close to the pericenter and provide orbital fits with and without the new data. In Section \ref{sec:high_e_degeneracy}, we describe a degeneracy that is inherent to the analysis of astrometric orbits of very eccentric visual binaries.  A discussion of the results in the context of the search for reliably very eccentric binaries concludes this paper in Section \ref{sec:discussion}. 

\section{Observations and new orbital solutions for $\zeta$ Boo and $\eta$ Oph}
\label{sec:results}

\subsection{Observations}
\label{sec:observations}

$\zeta$ Boo and $\eta$ Oph were observed with VLTI/GRAVITY \citep{GRAVITY17} in 2023 and 2024 throughout their pericenter passage. Unfortunately observations of $\zeta$ Boo during the passage itself were not possible because it was too close to the Sun. Tables \ref{table:obs_zeta_boo} and \ref{table:obs_eta_oph} summarize the observations including instrumental configurations and weather conditions. The observations were made with the four 1.8-m Auxiliary Telescopes (ATs). The baseline configuration, largest projected baseline $B_{\mathrm{proj,max}}$ and corresponding angular resolution $\theta_{\mathrm{max}}$ in the working near-infrared K band are also reported. In all nights the observations were made in single-field mode wherein half of the light is used to track the fringes at high temporal resolution and low spectral resolution while the other half is integrated at high spectral resolution (R=4,000) with 10s exposures. In each night between two to four files were obtained, each containing 16 exposures. The G8III star HD 129336 and the K2III star HD 153135, with angular diameters of 0.93 mas and 1.14 mas \citep{Bourges17}, were observed as interferometric calibrators for $\zeta$ Boo 
 and $\eta$ Oph respectively. The data were reduced with the ESO GRAVITY pipeline v.1.4.0 \citep{Lapeyrere14}. 

\begin{table*}
\centering
\caption{\label{table:obs_zeta_boo} Summary of the VLTI/GRAVITY observations of Zeta Bo\"otis.}
\begin{tabular}{ccccccc}
\hline \hline
Epoch & \shortstack{date\\MJD} & total exposure time & \shortstack{seeing\\@ 500 nm (")} & \shortstack{coherence time\\@ 500 nm (ms)} & \shortstack{AT configuration} & \shortstack{$B_{\mathrm{proj,max}}$ (m) \\$\theta_{\mathrm{max}}$ (mas)} \\ [0.3cm]

1 & \shortstack{2023-04-20\\60054.23} & 10s x 16 x 3 & 0.8-0.9 & 3 & K0-G2-D0-J3 & \shortstack{94.6\\4.8} \\ [0.3cm]

2 & \shortstack{2023-06-14\\60110.06} & 10s x 16 x 3 & 0.8-1.0 & 4 & K0-G2-D0-J3 & \shortstack{92.6\\4.9} \\ [0.3cm]

3 & \shortstack{2023-07-14\\60140.00} & 10s x 16 x 3 & 0.8-0.9 & 3 & K0-G2-D0-J3 & \shortstack{94.6\\4.8} \\ [0.3cm]

4 & \shortstack{2024-05-22\\60452.02} & 10s x 16 x 2 & 0.6 & 3 & A0-B2-D0-C1 & \shortstack{25.3\\18.0} \\ [0.3cm]

\hline
\end{tabular}
\end{table*}

\begin{table*}
\centering
\caption{\label{table:obs_eta_oph} Summary of the VLTI/GRAVITY observations of Eta Ophiuchi.}
\begin{tabular}{ccccccc}
\hline \hline
Epoch & \shortstack{date\\MJD} & total exposure time & \shortstack{seeing\\@ 500 nm (")} & \shortstack{coherence time\\@ 500 nm (ms)} & \shortstack{AT configuration} & \shortstack{$B_{\mathrm{proj,max}}$ (m) \\$\theta_{\mathrm{max}}$ (mas)} \\ [0.3cm]

1 & \shortstack{2023-06-16\\60111.15} & 10s x 16 x 4 & 0.50 & 3.5 & K0-G2-D0-J3 & \shortstack{102.4\\4.4} \\ [0.3cm]

2 & \shortstack{2024-05-18\\60448.24} & 10s x 16 x 2 & 1.1 & 2.1 & A0-G2-J2-J3 & \shortstack{130.5\\3.5} \\ [0.3cm]

3 & \shortstack{2024-06-17\\60478.25} & 10s x 16 x 2 & 0.96 & 2.3 & K0-B5-D0-J6 & \shortstack{201.8\\2.2} \\ [0.3cm]

4 & \shortstack{2024-07-08\\60499.21} & 10s x 16 x 2 & 0.9-1.2 & 2.6-3.7 & A0-B5-J2-J6 & \shortstack{201.5\\2.2} \\ [0.3cm]

5 & \shortstack{2024-07-26\\60517.09} & 10s x 16 x 2 & 0.85 & 2.6 & A0-B2-D0-C1 & \shortstack{33.8\\13.4} \\ [0.3cm]

6 & \shortstack{2024-08-11\\60533.05} & 10s x 16 x 2 & 0.6-0.8 & 2.0 & K0-G1-D0-J3 & \shortstack{131.4\\3.5} \\ [0.3cm]

\hline
\end{tabular}
\end{table*}

We note that for $\zeta$ Boo the default closure phases (which are the average over the 16 exposures in each file) had very high noise. Our investigation showed that this happened because the nearly equal brightness binary caused the fringe-tracker fiber to constantly move between the two peaks in the resulting acquisition camera image during the integration. In order to circumvent this problem, we inspected individual exposures in each file and found that one to three exposures in each file were stable and resulted in good data. We therefore worked with these individual good exposures rather than the averaged files in the case of $\zeta$ Boo. This is the reason the statistical errors in the interferometric data are somewhat larger for $\zeta$ Boo than for $\eta$ Oph when comparing Figures \ref{fig:gravity_fits_zeta_boo} and \ref{fig:gravity_fits_eta_oph}.

\subsection{Orbital separations from VLTI/GRAVITY}
\label{subsec:gravity_separations}

\begin{figure*}[]
\centering
\includegraphics[width=\textwidth]{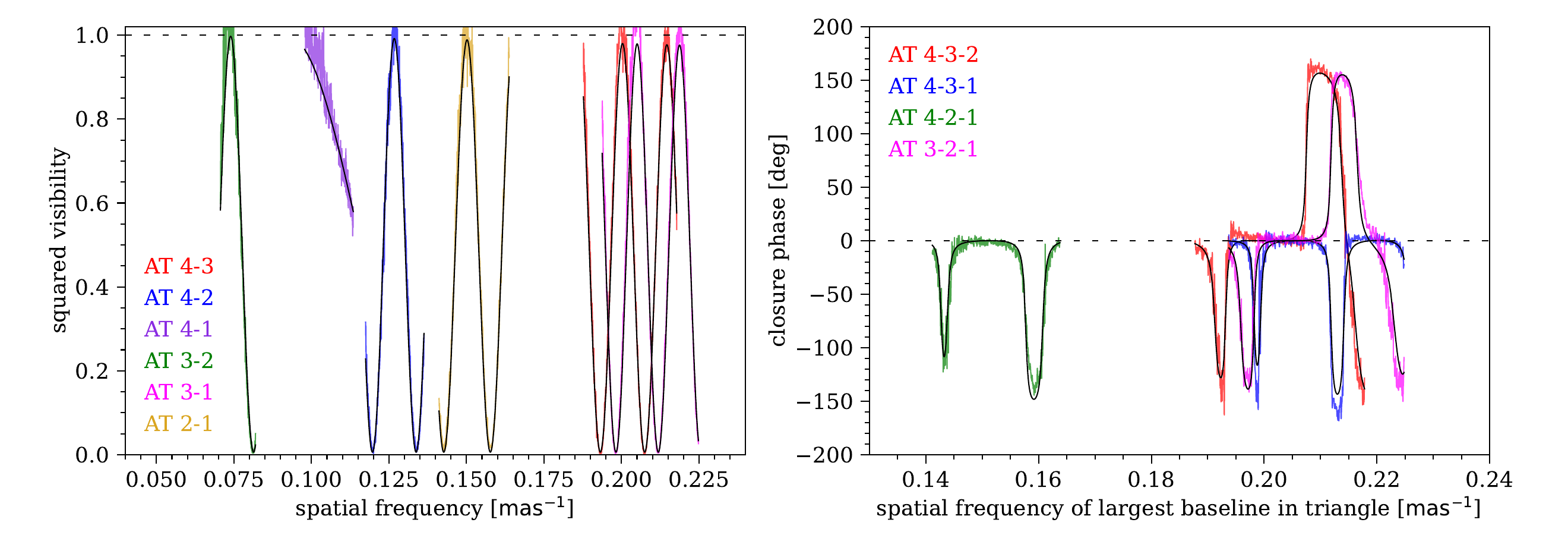} \\
\includegraphics[width=\textwidth]{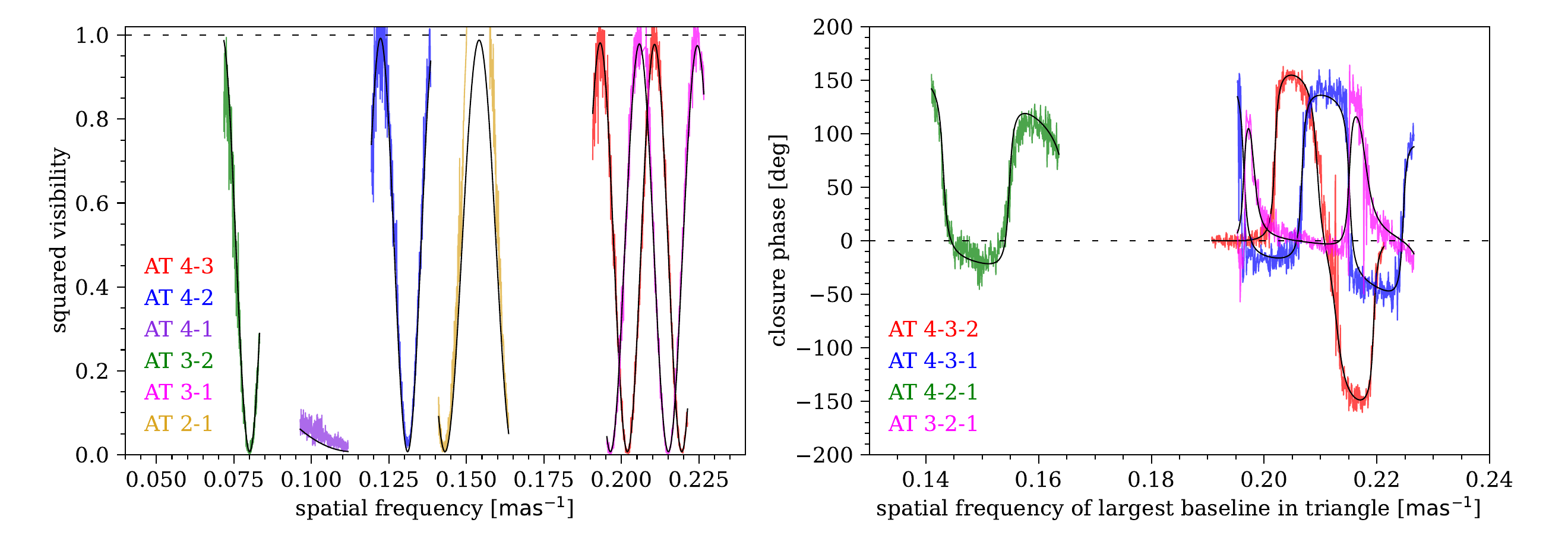} \\
\caption{\label{fig:gravity_fits_zeta_boo} VLTI/GRAVITY data (colored) and best fit binary model (solid black) for one example exposure in Epochs 1 (top) and 3 (bottom) of Zeta Bo\"otis. Each color corresponds to one baseline (left) or one triangle (right). The dashed lines show the expected signal for an unresolved single star.}
\end{figure*}

\begin{figure*}[]
\centering
\includegraphics[width=\textwidth]{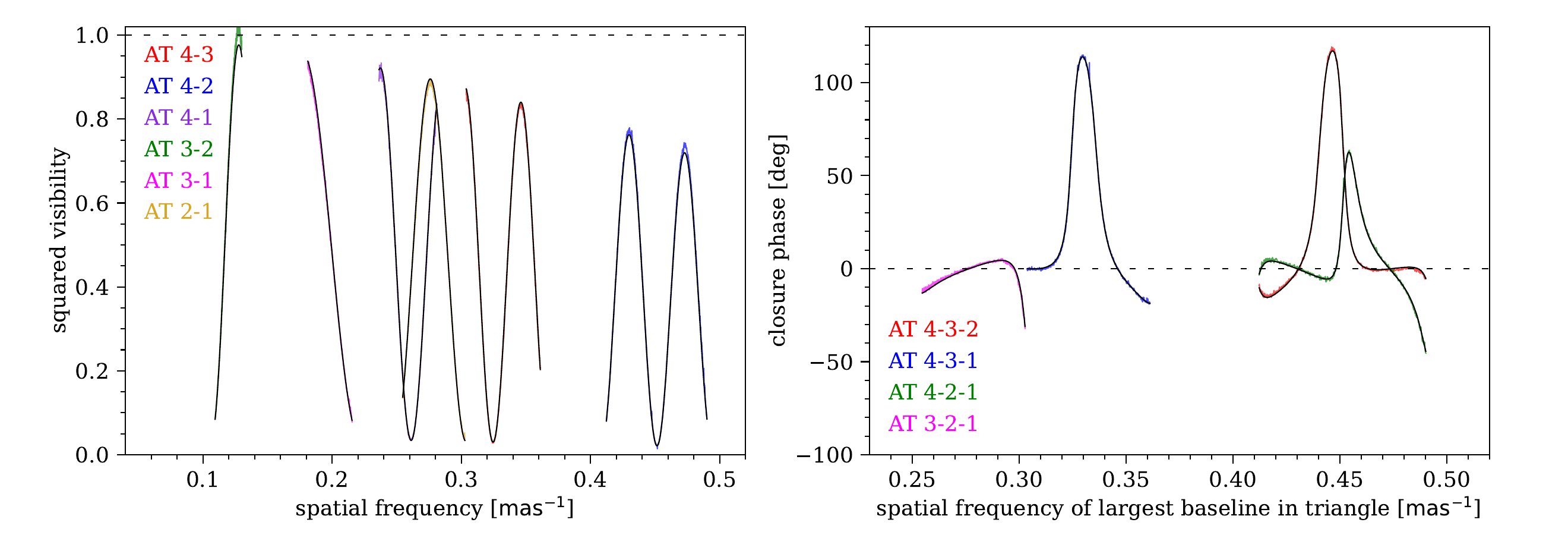}\\
\includegraphics[width=\textwidth]{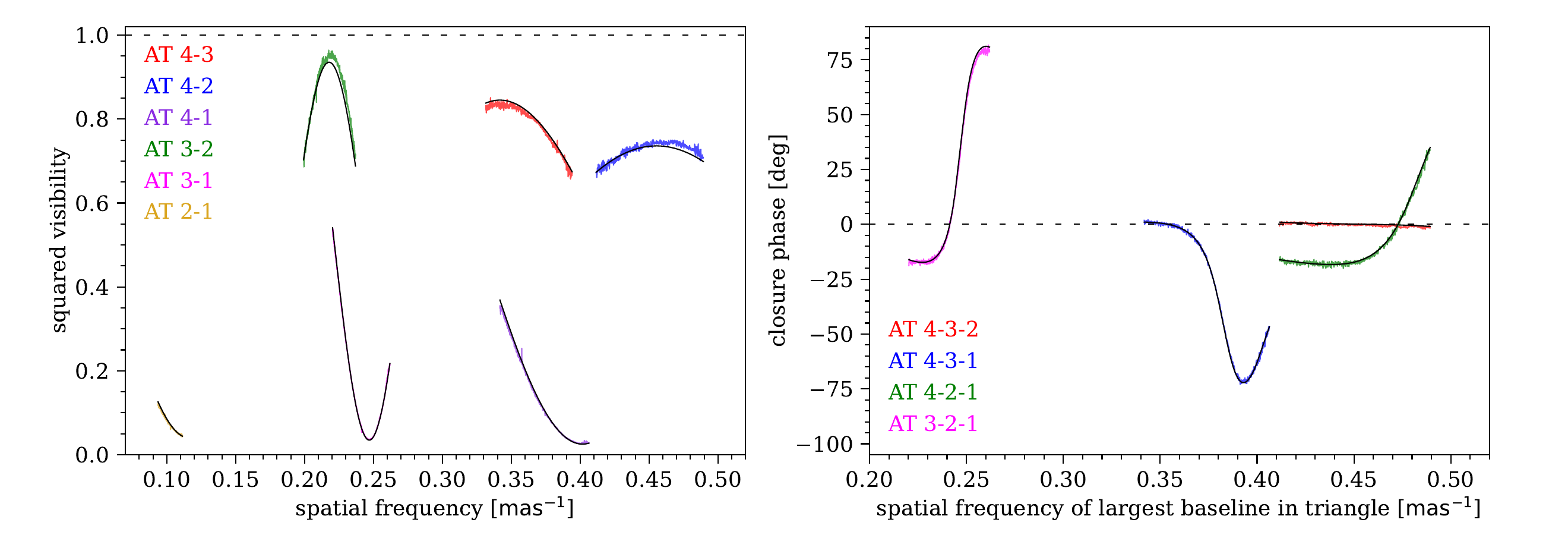}
\caption{\label{fig:gravity_fits_eta_oph} VLTI/GRAVITY data (colored) and best fit binary model (solid black) for one example file for Epochs 3 (top) and 4 (bottom) of Eta Ophiuchi. Each color corresponds to one baseline (left) or one triangle (right). The dashed lines show the expected signal for an unresolved single star.}
\end{figure*}

The interferometric data very clearly resolve the binary in both cases. We fit the data (namely, the squared visiblities and closure phases) with the interferometric binary model detailed in \cite{Waisberg23a}. The free parameters are the the K band flux ratio $\frac{f_B}{f_A}$ and the projected separation $(\Delta \alpha_*, \Delta \delta)_{B,A}$ of the secondary relative to the primary (towards the East and North directions on the sky plane). In addition, for two of the epochs of $\eta$ Oph with the highest angular resolution achieved with the extended AT array, the angular diameters of both components were also fit. The best-fit values, namely $\theta_{A} = 0.82 \pm 0.02 \text{ mas} \leftrightarrow 4.55 R_{\odot} $ and $\theta_{B} = 0.64 \pm 0.01 \text{ mas} \leftrightarrow 3.60 R_{\odot}$, were then fixed for the remaining epochs, in which these angular diameters are much below the interferometric resolution and cannot be well constrained. In the case of $\zeta$ Boo, the angular diameters were fixed to $\theta_A = 0.47 \text{ mas}$ and $\theta_B = 0.43 \text{ mas}$ based on the isochrone fitting described in \ref{subsec:isochrones} and have a negligible effect on the fit since they are well below the interferometric resolution for all the epochs. The best fit parameters for each epoch are reported in Tables \ref{table:gravity_results_zeta_boo} and \ref{table:gravity_results_eta_oph}. The interferometric data and best fit binary model are shown in Figures \ref{fig:gravity_fits_zeta_boo} and \ref{fig:gravity_fits_eta_oph} for two of the epochs for each binary. The corresponding figures for all remaining epochs are shown in Appendix \ref{app:additional_fits}. 

\begin{table}
\centering
\caption{\label{table:gravity_results_zeta_boo} Interferometric binary best fit results for Zeta Bo\"otis.}
\begin{tabular}{ccccc}
\hline \hline
Epoch & \shortstack{$\frac{f_{\mathrm{B}}}{f_{\mathrm{A}}}$ (\%)\\K band} & \shortstack{$\Delta \alpha_*$\\(mas)} & \shortstack{$\Delta \delta$\\(mas)} \\ [0.3cm]

1 & $90.5$ & $-67.42$ & $-31.14$ \\ [0.3cm]

2 & $88.1$ & $-53.92$ & $-33.40$ \\ [0.3cm]

3 & $85.3$ & $-45.98$ & $-34.43$ \\ [0.3cm]

4 & $89.3$ & $-17.30$ & $86.10$ \\ [0.3cm]

\hline 

average & $88.3 \pm 1.9$ & - & - \\ [0.3cm]

\hline
\end{tabular}
\end{table}

\begin{table}
\centering
\caption{\label{table:gravity_results_eta_oph} Interferometric binary fit results for Eta Ophiuchi.}
\begin{tabular}{cccc}
\hline \hline
Epoch & \shortstack{$\frac{f_{\mathrm{B}}}{f_{\mathrm{A}}}$ (\%)\\K band} & \shortstack{$\Delta \alpha_*$\\(mas)} & \shortstack{$\Delta \delta$\\(mas)} \\ [0.3cm]

1 & $50$\tablenotemark{a} & $-176.36$ & $-197.36$ \\ [0.3cm]

2 & $63.5$ & $-29.46$ & $-49.36$ \\ [0.3cm]

3 & $66.1$ & $-6.62$ & $-22.54$ \\ [0.3cm]

4 & $66.3$ & $9.82$ & $-2.46$ \\ [0.3cm]

5 & $63.6$ & $23.40$ & $14.80$ \\ [0.3cm]

6 & $67.5$ & $34.97$ & $29.77$ \\ [0.3cm]

\hline 

average & $65.4 \pm 1.6$ & - & - \\ [0.3cm]

\hline
\end{tabular}
\tablenotetext{1}{Very uncertain due to the very large separation compared to the GRAVITY fiber FOV and not included in the average.}
\end{table}

The averaged K band flux ratio and the error estimated from the scatter between the different epochs are reported at the end of the tables. For the separations, the formal fit errors are typically below one micro-arcsecond and are expected to be very underestimated due to data correlation and systematic errors. In order to estimate the errors we instead look at the average scatter between the best fit results for different files/exposures across a given epoch. The estimated errors are $40$ and $20$ $\mu$as for $\zeta$ Boo and $\eta$ Oph respectively (Table \ref{table:estimated_errors}). In order to check whether these errors are reasonable, we performed an orbital fit on the GRAVITY data alone. Despite the parameters being uncertain and degenerate due to the very small orbital coverage, the fit residuals are comparable with the estimated errors, suggesting that the latter are reasonable.

In both cases, there is no evidence for any additional companion in the interferometric data. Following the method outlined in Section 4.2.3 in \cite{Waisberg23a}, we can exclude close companions to either component of
$\zeta$ Boo within $\rho \gtrsim 3 \text{ mas}$ with a K band flux ratio above $1\%$ (corresponding to $K_{\mathrm{abs}} = 5.8$ or a $0.45 M_{\odot}$ M dwarf) and within $1 \lesssim \rho \lesssim 3 \text{ mas}$ with a K band flux ratio above 2$\%$ (corresponding to $K_{\mathrm{abs}} \approx 5.1$ or a $0.6 M_{\odot}$ M dwarf). For $\eta$ Oph, we can exclude close companions with a K band flux ratio larger than $0.5\%$, corresponding to $K_{\mathrm{abs}}= 6.6$ and 7.0 for the primary and secondary respectively (about $0.3-0.4 M_{\odot}$ M dwarfs). 

\subsection{Isochrone masses and radii}
\label{subsec:isochrones}

We estimated the age and individual component masses in each system by fitting Mesa Isochrones and Stellar Tracks \citep[\texttt{MIST}; ][]{Dotter16,Choi16,Paxton11,Paxton13,Paxton15} packaged model grids\footnote{\url{https://waps.cfa.harvard.edu/MIST/model_grids.html}} to the total K band magnitude from 2MASS \citep[$K=3.697\pm0.208$ for $\zeta$ Boo and $K=2.336\pm0.242$ for $\eta$ Oph;][]{Skrutskie06}, the K band interferometric flux ratio from the VLTI/GRAVITY interferometric observations and the individual Tycho $V_T$ and $B_T$ magnitudes from \cite{Fabricius00} ($V_{T,A} = 4.53$, $V_{T,B} = 4.57$, $B_{T,A} = 4.60$ and $B_{T,B} = 4.66$ for $\zeta$ Boo; and $V_{T,A} = 2.97$, $V_{T,B} = 3.44$, $B_{T,A} = 3.10$ and $B_{T,B} = 3.56$ for $\eta$ Oph). The statistical error is dominated by the relatively large 2MASS photometric error for our very bright targets. We follow the same methodology as outlined in Section 4.4 of \cite{Waisberg23a}. We note that the interstellar extinction is negligible for both targets. We adopted the isochrones with solar metallicity and rotational velocity $\frac{v}{v_{\mathrm{crit}}}=0.4$. For $\eta$ Oph, $v \sin i = 10 \text{ km}\text{ s}^{-1}$ so that if the rotational axes are aligned with the orbital axis its intrinsic rotational velocity is very low. If we adopt the isochrones with zero rotational velocity instead then the stars would be slightly older by about 60 Myr and slightly less massive by about $0.03 M_{\odot}$. However, these differences are of the order or below the resulting uncertainties in any case. 

Table \ref{table:isochrone} reports the isochrone fitting results. For $\zeta$ Boo we find a total mass $M_{A+B} =4.36_{-0.08}^{+0.24} M_{\odot}$. The isochrone angular diameters $\theta_A = 0.47 \pm 0.07 \text{ mas}$ and $\theta_B = 0.43 \pm 0.05 \text{ mas}$ were used to fix the parameters in the interferometric fitting since they cannot be directly measured from the current interferometric data.

For $\eta$ Oph we find a total isochrone mass $M_{A+B} = 4.28_{-0.11}^{+0.20} M_{\odot}$. The isochrone radii translate to angular diameters $\theta_A = 0.90 \pm 0.14 \text{ mas}$ and $\theta_B = 0.72 \pm 0.11 \text{ mas}$, which are consistent with the interferometric measurements within 1$\sigma$.

\begin{table}
\centering
\caption{\label{table:isochrone} Age, masses, temperatures and radii estimates for the components based on MIST isochrones.}
\begin{tabular}{ccc}
\hline \hline
& $\zeta$ Boo 
& $\eta$ Oph 
\\ [0.3cm]

\shortstack{age\\(Myr)} 
& $560_{-240}^{+150}$ 
& $500_{-280}^{+210}$ 
\\ [0.3cm]

\shortstack{mass\\($M_{\odot}$)} 
& \shortstack{$M_A=2.21_{-0.05}^{+0.14}$\\$M_B=2.15_{-0.03}^{+0.10}$} 
& \shortstack{$M_A=2.25_{-0.09}^{+0.18}$\\$M_B=2.03_{-0.02}^{+0.02}$} 
\\ [0.3cm]

\shortstack{$T_{\mathrm{eff}}$\\(K)} 
& \shortstack{$T_{\mathrm{eff,A}}=8800_{-600}^{+1000}$\\$T_{\mathrm{eff,A}}=8750_{-550}^{+800}$} 
& \shortstack{$T_{\mathrm{eff,A}}=9000_{-750}^{+1250}$\\$T_{\mathrm{eff,A}}=8700_{-450}^{+450}$} 
\\ [0.3cm]

\shortstack{$R$\\($R_{\odot}$)} 
& \shortstack{$R_A=2.6_{-0.4}^{+0.3}$\\$R_B=2.4_{-0.3}^{+0.3}$} 
& \shortstack{$R_A=2.5_{-0.4}^{+0.4}$\\$R_B=2.0_{-0.2}^{+0.3}$} 
\\ [0.3cm]

\hline
\end{tabular}
\end{table}

We stress that the isochrone parameters may have larger uncertainties due to systematic errors from the stellar evolution models and additional assumptions that may not completely hold (for example that the components have evolved completely independently since birth). Model-independent measurements (such as the interferometric radii in the case of $\eta$ Oph and the dynamical masses from the orbital fitting below) may in principle be used to constrain parameters in the evolutionary models but this is beyond the scope of this paper.

\subsection{Orbital fit}
\label{subsec:orbital_fit}

The astrometric data for $\zeta$ Boo and $\eta$ Oph in ORB6 are very heterogeneous, spanning the years 1796 to 2022 and 1889 to 2022 respectively, as well as a large number of techniques and observers. We divided the ORB6 data into groups of low resolution and high resolution observations in order to estimate the uncertainties and exclude outliers. In Appendix \ref{app:orb6_details} we provide details on the data selection and error estimation for the ORB6 data. The number of astrometric points and the estimated errors for each group are reported in Table \ref{table:estimated_errors}. An orbital fit can then be performed by combining the different groups with their appropriate errors. 

\begin{table}
\centering
\caption{\label{table:estimated_errors} Estimated astrometric errors for each data group.}
\begin{tabular}{ccccc}
\hline \hline
& VLTI/GRAVITY & \shortstack{ORB6\\PTI} & \shortstack{ORB6\\low res.} & \shortstack{ORB6\\high res.} \\ [0.3cm]

$\zeta$ Boo & \shortstack{40 $\mu$as\\ (4)} & \shortstack{2.4 mas\\ (17)} & \shortstack{19.3 mas \\ (125)} & \shortstack{78.4 mas \\ (615)} \\ [0.3cm]

$\eta$ Oph & \shortstack{20 $\mu$as\\ (6)} & - & \shortstack{13.0 mas \\ (106)} & \shortstack{53.5 mas \\ (137)} \\ [0.3cm]

\hline
\end{tabular}
\tablenotetext{0}{The second line in each entry reports the number of astrometric points after excluding outliers.}
\end{table}

We fit for the seven Keplerian elements\footnote{We note that due to the very high eccentricity of $\zeta$ Boo, it was necessary to solve Kepler's Equation for the eccentric anomaly (E) to very high accuracy in order to find the best fit orbital solution. We find that a tolerance $\mathrm{E} - e \sin \mathrm{E} - \mathrm{MA} < 10^{-8}$ (where MA is the mean anomaly) was needed so we adopted a tolerance of $10^{-9}$.} (angular semi-major axis $a$, eccentricity $e$, inclination $i$, longitude of the ascending node $\Omega$, argument of periastron $\omega$, period $P$ and time of periastron $T_p$) using nonlinear least squares optimization implemented with the \texttt{lmfit}\footnote{\url{https://pypi.org/project/lmfit/}} package. The parameter uncertainties (defined as the 2.3\% and 97.7\% percentiles in the distribution) were estimated by generating and fitting $2\times10^3$ random resamples of the astrometric data according to their estimated errors to get the distribution of the best fit parameters. 

Table \ref{table:orbital_fit_results} reports the best fit results for both binaries for two cases, namely without and with the VLTI/GRAVITY data throughout pericenter passage \footnote{Note that without RV curves there is a degeneracy between the two solutions with $(\Omega, \omega)$ and $(\Omega + 180^{\degr}, \omega + 180^{\degr})$. We report the solution with $\Omega < 180^{\degr}$ as is standard practice.}. The resulting physical semi-major axes, pericenter distances and the dynamical masses are calculated using the distances adopted in Section \ref{sec:introduction}. 

\begin{table*}
\centering
\caption{\label{table:orbital_fit_results} Best fit Keplerian parameters for the orbits of Zeta Bo\"otis and Eta Ophiuchi. The values refer to the best fit orbit and the uncertainties correspond to the 2.3\% and 97.7\% percentiles of the parameter distributions.}
\begin{tabular}{ccccc}
\hline \hline
& \shortstack{$\zeta$ Boo\\ORB6} & \shortstack{$\zeta$ Boo\\ORB6+GRAVITY} 
& \shortstack{$\eta$ Oph\\ORB6} & \shortstack{$\eta$ Oph\\ORB6+GRAVITY}  
\\ [0.3cm]

\shortstack{a\\(mas)\\(au)} 
& \shortstack{$1476.7_{-220.4}^{+500.9}$\\$76.4_{-11.5}^{+26.0}$} & \shortstack{$808.6_{-2.0}^{+1.9}$\\$41.84_{-0.44}^{+0.46}$} 
& \shortstack{$1144.9_{-137.1}^{+234.8}$\\$29.5_{-3.7}^{+6.5}$} & \shortstack{$1203.8_{-2.6}^{+3.0}$\\$31.05_{-1.35}^{+1.41}$} 
\\ [0.3cm]

e 
& $0.9942_{-0.0023}^{+0.0026}$ & $0.980450_{-0.000064}^{+0.000063}$ 
& $0.924_{-0.023}^{+0.024}$ & $0.93077_{-0.00012}^{+0.00013}$ 
\\ [0.3cm]

\shortstack{$i$\\(deg)} 
& $119.5_{-5.4}^{+3.5}$ & $125.88_{-0.16}^{+0.16}$ 
& $96.5_{-1.2}^{+1.0}$ & $96.305_{-0.018}^{+0.019}$ 
\\ [0.3cm]

\shortstack{$\Omega$\\(deg)} 
& $187.0_{-3.5}^{+3.1}$ & $176.63_{-0.16}^{+0.16}$ 
& $40.10_{-0.57}^{+0.59}$ & $39.241_{-0.011}^{+0.011}$ 
\\ [0.3cm]

\shortstack{$\omega$\\(deg)} 
& $78.2_{-2.9}^{+3.5}$ & $62.08_{-0.14}^{+0.13}$ 
& $276.4_{-1.2}^{+1.1}$ & $275.90_{-0.11}^{+0.10}$ 
\\ [0.3cm]

\shortstack{P\\(yrs)} 
& $123.97_{-0.46}^{+0.42}$ & $125.04_{-0.21}^{+0.24}$ 
& $87.68_{-0.52}^{+0.53}$ & $87.77_{-0.24}^{+0.28}$ 
\\ [0.3cm]

$T_p$ 
& $2023.26_{-0.15}^{+0.50}$ & $2023.9548_{-0.0014}^{+0.0013}$
& $2024.42_{-0.44}^{+0.43}$ & $2024.52608_{-0.00031}^{+0.00029}$ 
\\ [0.3cm]

\hline \\ [0.1cm]

\shortstack{$a_p$\\(au)} 
& $0.44_{-0.12}^{+0.09}$ & $0.818_{-0.008}^{+0.009}$ 
& $2.2_{-0.4}^{+0.4}$ & $2.15_{-0.09}^{+0.10}$ 
\\ [0.3cm]

\shortstack{$M_{\mathrm{dyn}}$\\($M_{\odot}$)} 
& $29.0_{-11.3}^{+41.1}$ & $4.69_{-0.15}^{+0.15}$ 
& $3.4_{-1.2}^{+2.6}$ & $3.89_{-0.49}^{+0.56}$ 
\\ [0.3cm]

\hline \\ [0.1cm]

\shortstack{$M_{xx}$\\(mas)} 
& $-359.0_{-0.1}^{+4.7}$ & $-353.35_{-0.93}^{+0.99}$ 
& $14.2_{-2.4}^{+2.4}$ & $12.57_{-1.54}^{+1.45}$ 
\\ [0.3cm]

\shortstack{$M_{xy}$\\(mas)} 
& $152.4_{-5.5}^{+6.1}$ & $142.92_{-0.24}^{+0.23}$  
& $335.3_{-4.3}^{+4.2}$ & $342.19_{-0.70}^{+0.75}$ 
\\ [0.3cm]

\shortstack{$M_{yx}$\\(mas)} 
& $443.1_{-3.9}^{+0.7}$ & $440.36_{-0.79}^{+0.82}$
& $179.8_{-4.2}^{+4.0}$ & $180.04_{-1.29}^{+1.28}$ 
\\ [0.3cm]

\shortstack{$M_{yy}$\\(mas)} 
& $29.5_{-9.5}^{+5.2}$ & $35.33_{-0.30}^{+0.30}$ 
& $275.1_{-4.9}^{+4.9}$ & $273.08_{-0.68}^{+0.70}$ 
\\ [0.3cm]

\hline
\end{tabular}
\tablenotetext{0}{For the physical distances and dynamical masses, the adopted distances are 
$d=51.7\pm0.2 \text{ pc}$ and  $d=25.8\pm0.6 \text{ pc}$.}
\end{table*}

\begin{figure*}[]
 \includegraphics[width=2\columnwidth]{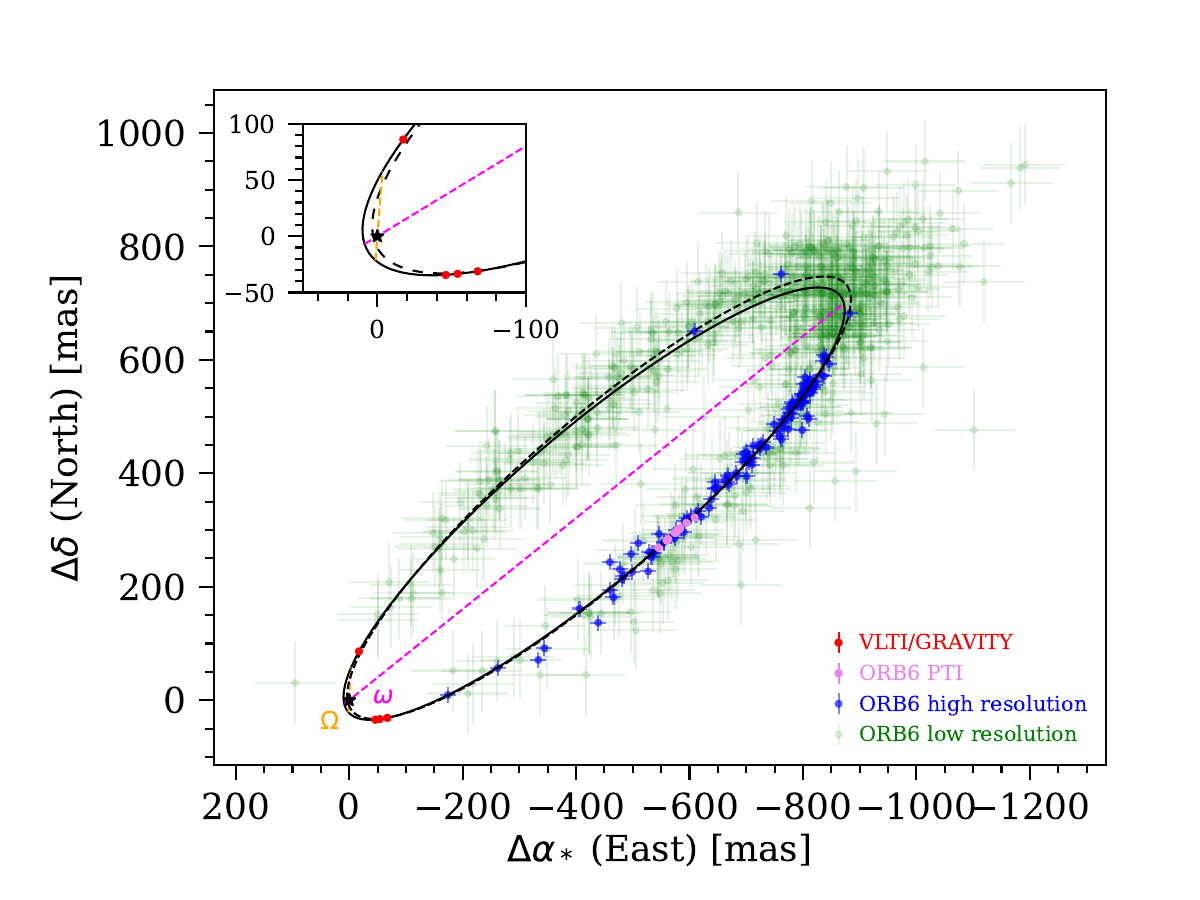}
 \caption{ \label{fig:fit_orbit_zeta_boo} Astrometric data (red, violet, blue, green) and best fit orbit (solid black) for Zeta Bo\"otis. The best fit orbit without including the VLTI/GRAVITY points is shown in a black dashed line. The zoomed inset shows the 2023/2024 periastron passage. The black star marks the position of the primary. The line of apsides ($\omega$) and the line of nodes ($\Omega$) are shown in magenta and orange.}
\end{figure*}

\begin{figure*}[]
 \includegraphics[width=2\columnwidth]{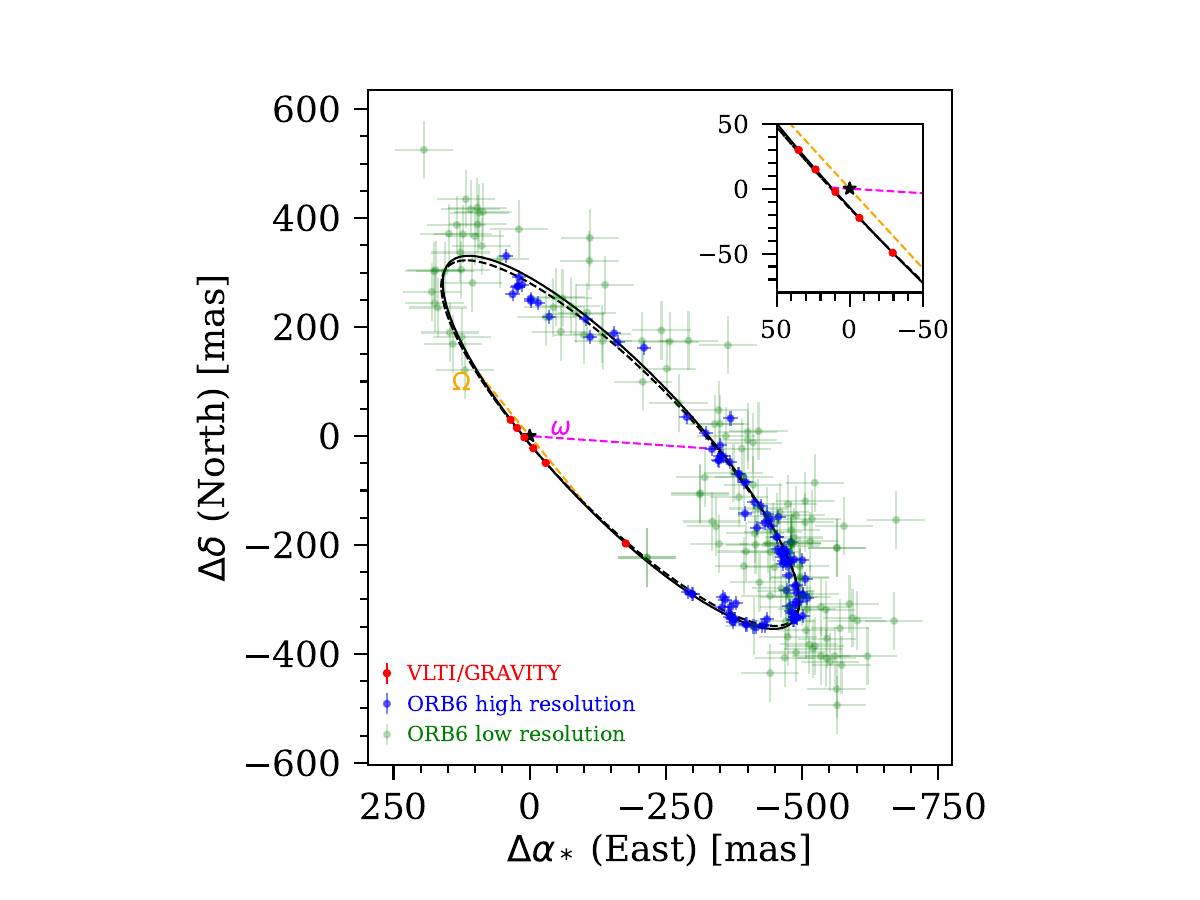}
 \caption{\label{fig:fit_orbit_eta_oph} Astrometric data (red, blue, green) and best fit orbit (solid black) for Eta Ophiuchi. The best fit orbit without including the VLTI/GRAVITY points is shown in a black dashed line. The zoomed inset shows the 2024 periastron passage. The black star marks the position of the primary. The line of apsides ($\omega$) and the line of nodes ($\Omega$) are shown in magenta and orange.}
\end{figure*}

\begin{figure}[]
 \includegraphics[width=1\columnwidth]{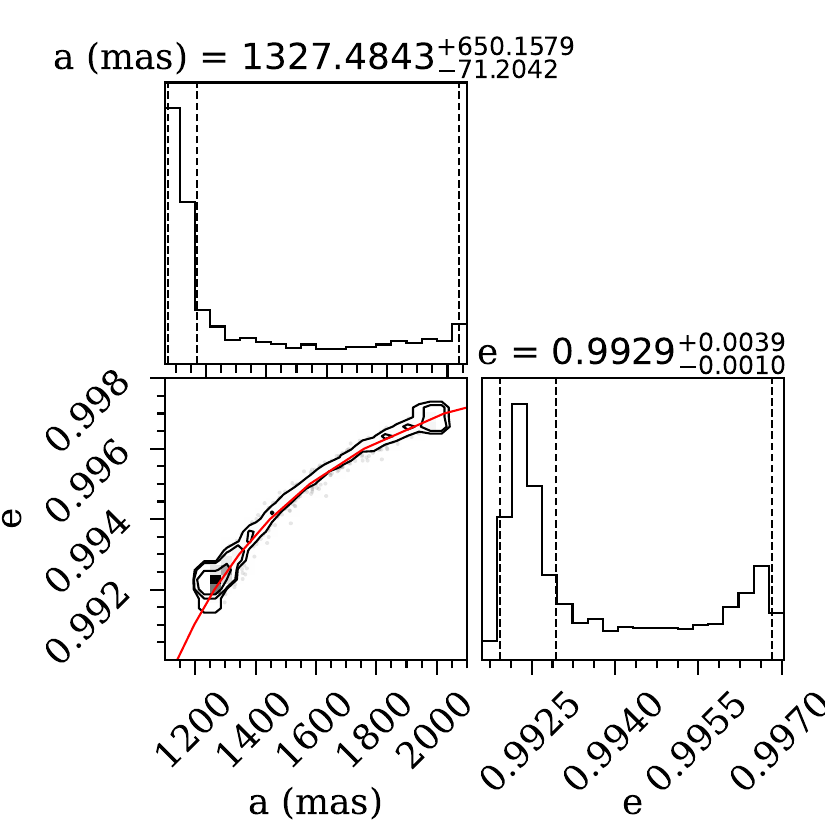} \\
 \includegraphics[width=1\columnwidth]{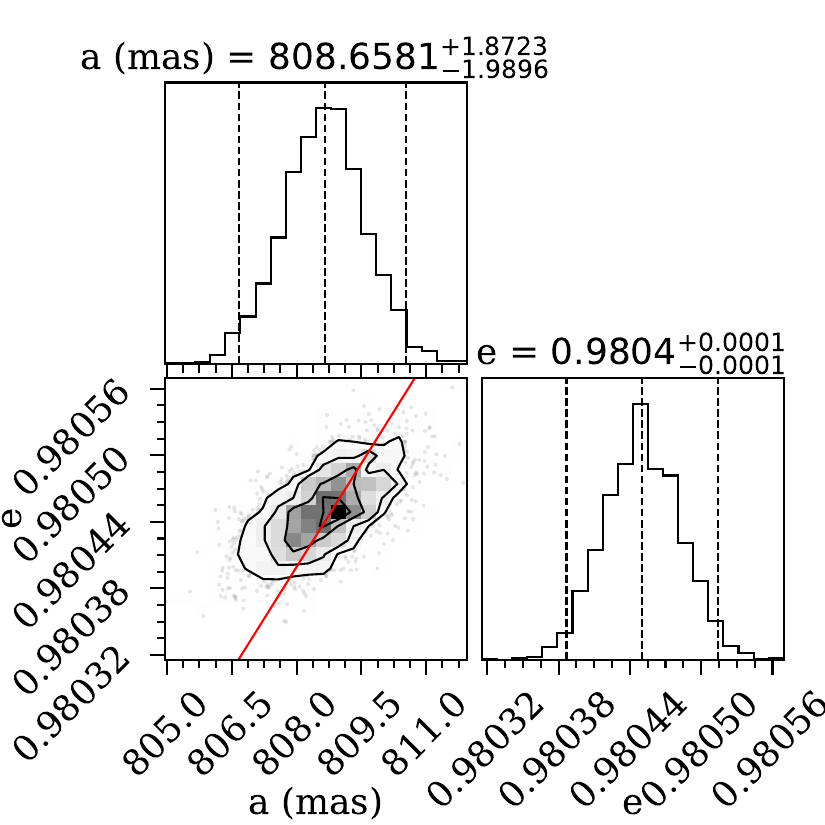}
 \caption{\label{fig:a_e_degeneracy_zeta_Boo} Orbital fit distributions of semi-major axis $a$ and eccentricity $e$ for Zeta Bo\"otis without (top) and with (bottom) including the VLTI/GRAVITY data covering the pericenter passage. The dashed lines and quoted numbers refer to the 2.3\%, 50\% and 97.7\% quantiles. The red lines show the predicted $a-e$ degeneracy line for a very eccentric visual binary without pericenter coverage (Eq. \ref{eq:eadegeneracy}). }
\end{figure}

\begin{figure}[]
 \includegraphics[width=1\columnwidth]{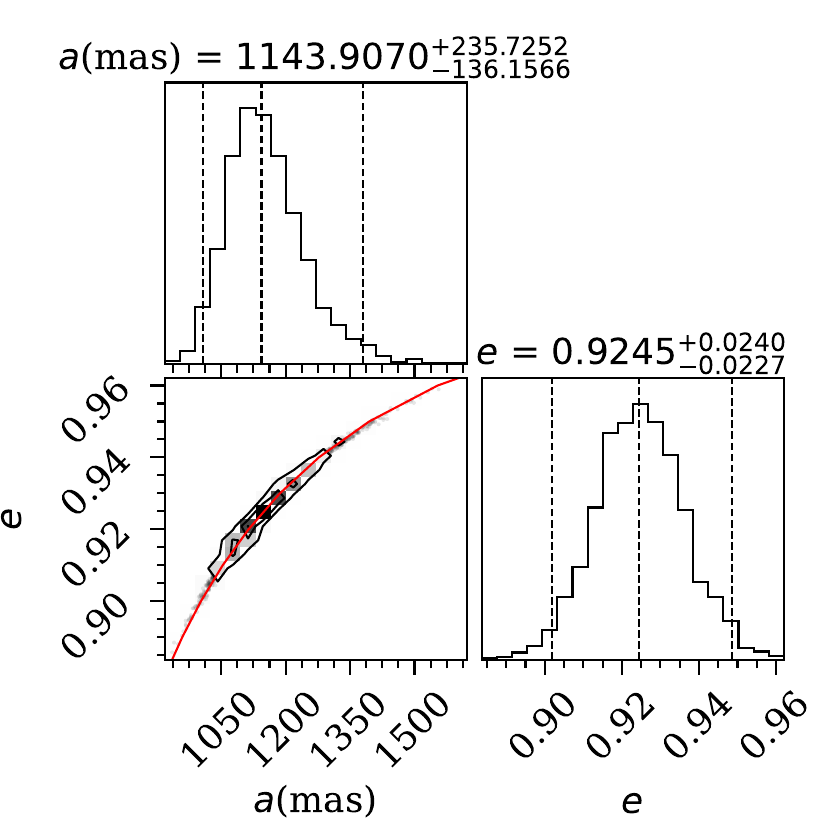} \\
 \includegraphics[width=1\columnwidth]{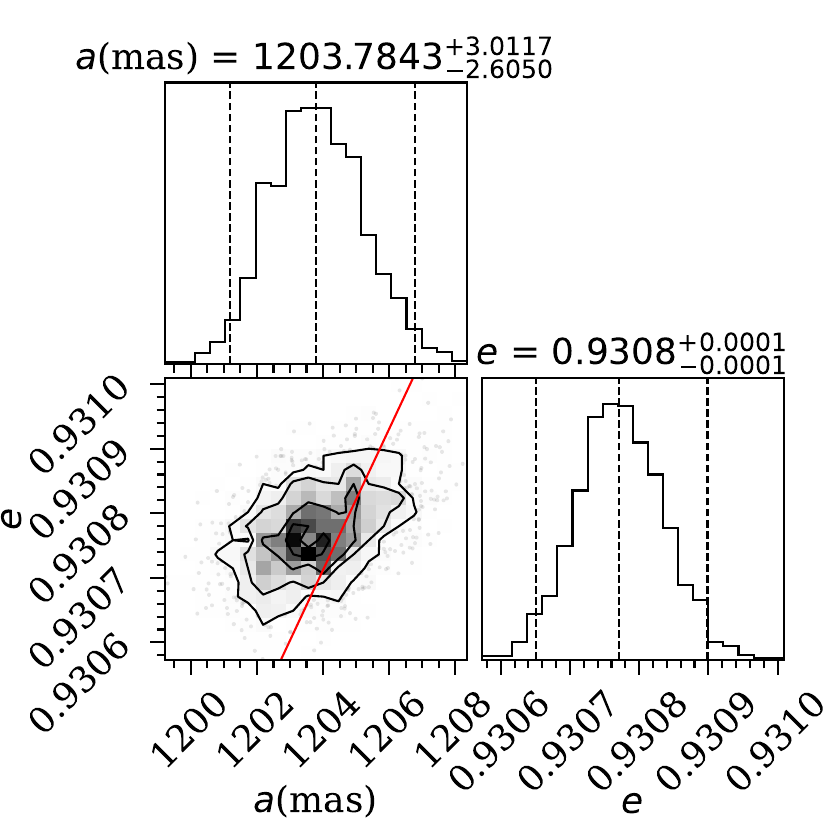}
 \caption{\label{fig:a_e_degeneracy_eta_oph} Same as figure \ref{fig:a_e_degeneracy_eta_oph} but for Eta Ophiuchi.}
\end{figure}

Figures \ref{fig:fit_orbit_zeta_boo} and \ref{fig:fit_orbit_eta_oph} display the data and the best fit orbits for $\zeta$ Boo and $\eta$ Oph respectively, with a zoom inset of the periastron passage. The solid black line shows the best fit orbits for the combined ORB6 and VLTI/GRAVITY data, while the dashed lines show the best fit orbits for the ORB6 data alone. The line of apisdes ($\omega$) is shown in a dashed magenta line and the line of nodes ($\Omega$) is shown in a dashed orange line. According to the best fit combined solution, the periastron passages occurred on 2023 December 16th and 2024 July 12th with uncertainties of about 11 hrs and 2.5 hrs for $\zeta$ Boo and $\eta$ Oph respectively. Figures \ref{fig:a_e_degeneracy_zeta_Boo} and \ref{fig:a_e_degeneracy_eta_oph} show the resulting $a-e$ distribution for the orbital fits without and with including the VLTI/GRAVITY pericenter data for $\zeta$ Boo and $\eta$ Oph respectively and will be discussed in detail in Section \ref{sec:high_e_degeneracy} below. Note that the $a$ and $e$ values quoted in Figures \ref{fig:a_e_degeneracy_zeta_Boo} and \ref{fig:a_e_degeneracy_eta_oph} correspond to the 2.3, 50 and 97.7\% quantiles of the resulting distribution so that they do not exactly correspond to the values in Table \ref{table:orbital_fit_results} and which refer to the best-fit orbit.

In the case of $\zeta$ Boo, the orbital fit without the VLTI/GRAVITY pericenter data has not only higher uncertainties but is also biased towards large eccentricity values $e \gtrsim 0.99$ \citep[consistent with the solution in][]{Muterspaugh10a}. The solution is clearly unphysical as it results in an unreasonably high dynamical mass $M_{\mathrm{dyn}} = 29.0_{-11.3}^{+41.1} M_{\odot}$. Somewhat confusingly, the ``correct'' solution, as revealed in the orbital fit including the VLTI/GRAVITY pericenter data, is excluded by the fit at high significance. Figure \ref{fig:a_e_degeneracy_zeta_Boo} shows that without the VLTI/GRAVITY pericenter data there is a very strong correlation between $a$ and $e$ with a long tail towards very high eccentricities, hinting at the likely culprit of this problematic solution. 
 
Adding the VLTI/GRAVITY pericenter data results in a much better behaved solution with a high but not so extreme eccentricity $e=0.98045 \pm 0.00006$ and a smaller semi-major axis $a=808.6_{-2.0}^{+1.9} \text{ mas}$, which imply a pericenter distance $a_p = 0.818 \pm 0.008 \text{ au}$ and a reasonable dynamical mass $M_{\mathrm{dyn}} = 4.69 \pm 0.15 M_{\odot}$ which is consistent with the photometric mass inferred from the isochrones within their errors. As Figure \ref{fig:a_e_degeneracy_zeta_Boo} shows, adding the VLTI/GRAVITY pericenter data breaks the strong correlation between $a$ and $e$ that was present when using the ORB6 data alone. 

In the case of $\eta$ Oph, we found that the data without VLTI/GRAVITY constrain $e=0.924 \pm 0.024$ and $a=1145_{-137}^{+235} \text{ mas}$. The eccentricity is consistent with the $e=0.95 \pm 0.02$ solution currently listed in ORB6 but we find that the latter's uncertainty in semi-major axis of 10 mas is highly underestimated. Because of the large error in $a$ the pericenter distance and dynamical mass are only poorly constrained. Adding the VLTI/GRAVITY pericenter data leads to much tighter constraints $e=0.93077\pm0.00013$ and $a=1203.8_{-2.6}^{+3.0} \text{ mas}$, resulting in a pericenter distance $a_p=2.15 \pm 0.10 \text{ au}$ and a dynamical mass $M_{\mathrm{dyn}} = 3.89 \pm 0.53 M_{\odot}$ (the larger errors relative to $\zeta$ Boo are due to the larger distance error). We therefore confirm that the high dynamical mass $M_{\mathrm{dyn}} \simeq 6 M_{\odot}$ of the current orbital solution in ORB6 is due to an underestimated error in $a$ rather than any unresolved close companion. As is the case for $\zeta$ Boo, Figure \ref{fig:a_e_degeneracy_eta_oph} shows that the solution with the ORB6 data alone has a strong degeneracy between $a$ and $e$ that is broken when adding the VLTI/GRAVITY pericenter data.

\section{Degeneracy in very eccentric visual orbits}
\label{sec:high_e_degeneracy}

As can be seen in the top panels of Figures \ref{fig:a_e_degeneracy_zeta_Boo} and \ref{fig:a_e_degeneracy_eta_oph}, there is a strong correlation between semi-major axis and eccentricity in the orbital solution for the ORB6 data that does not include the pericenter passage. Moreover, the best fit values obtained for $\zeta$ Boo from these data are apparently not consistent with the more accurate fit that includes the new data by a large margin. Our goal in this section is to clarify the origin of the degeneracy that is apparent here and which can plague the orbital solutions for very eccentric visual binaries.

When viewed face-on, orbits with varying values of eccentricities which are very close to unity can be easily distinguished due to their significantly different semi-minor to semi-major axes ratios, even when only data far away from the pericenter passage is available. However, the situation is less clear for other viewing angles given the projection of the ellipse on the sky. As we next show, far from the pericenter passage, a very eccentric orbit can have the same sky-projected orbit as other arbitrarily high eccentricities by suitable changes of the viewing angles and semi-major axis (see for e.g. the two orbital solutions for $\zeta$ Boo in Figure \ref{fig:fit_orbit_zeta_boo}).

In a frame aligned with the orbit, with the x-axis in the pericenter direction, the position of the secondary relative to the primary normalized by the semi-major axis is given by:

\begin{align}\label{eq:xoyo}
\begin{bmatrix}
\hat x_o \\
\hat y_o
\end{bmatrix}
=
\begin{bmatrix}
\cos(E) - e \\
\sqrt{1-e^2} \sin(E)
\end{bmatrix}
\end{align}
where $E$ is the eccentric anomaly given by Kepler's Equation 
\begin{align} \label{eq:Kepler}
2 \pi \frac{t - T_p}{P} = E - e \sin(E),
\end{align}
\noindent $T_p$ is the periastron epoch and $P$ is the orbital period.

The observed 2d positions on the sky depend further on the angular semi-major axis, sky orientation, and viewing angles, which all affect the sky position through 2d linear transformations of the normalized face-on coordinates. The final positions $x$ (with respect to the East) and $y$ (with respect to the North) can be expressed as 

\begin{align}\label{eq:xyMpxoyo}
\begin{bmatrix}
x\\
y
\end{bmatrix}
=
\mathbf{M'}\begin{bmatrix}
\hat x_o \\
\hat y_o
\end{bmatrix}
\end{align}
where $\mathbf{M'}$ is a 2x2 matrix that is related to $a,i,\Omega,\omega$ through
\noindent 
\begin{align}\label{eq:MpOmiom}
\mathbf{M'}=a\cdot \mathbf{R}(\Omega)\cdot\begin{bmatrix}
1 & 0 \\
0 & \cos i
\end{bmatrix}
\cdot \mathbf{R}(\omega),
\end{align}
where
\begin{align}
\mathbf{R}(\theta)=
\begin{bmatrix}
\cos \theta & -\sin \theta \\
\sin \theta & \cos \theta
\end{bmatrix}
\end{align}
is the 2D rotation matrix.
The matrix $\mathbf{M'}$ is often parametrized as 
\begin{align}
\mathbf{M'}=\begin{bmatrix}
M_{xx}' & M_{xy}' \\
M_{yx}' & M_{yy}'
\end{bmatrix}=\begin{bmatrix}
A & F \\
B & G
\end{bmatrix}
\end{align}
and its 4 elements $M_{ij}' = A,F,B,G$ (Thiele-Innes elements) can obtain any set of 4 values. They are often used for parameterizing visual orbits instead of the four parameters $a,\omega,\Omega,i$.

To study the limit of high eccentricities more carefully, it is useful to separate the factor $\sqrt{1-e^2}$ from the other effects of the eccentricity on the position in equations \eqref{eq:xoyo} and \eqref{eq:xyMpxoyo}:

\begin{align}\label{eq:xyM}
\begin{bmatrix}
x\\
y
\end{bmatrix}
=
\mathbf{M}\cdot 
\begin{bmatrix}
\cos(E) - e \\
\sin(E)
\end{bmatrix}, 
\end{align}
where
\begin{align}\label{eq:MM'}
\mathbf{M}=\mathbf{M'}\cdot \begin{bmatrix}
1 & 0 \\
0 & \sqrt{1-e^2}
\end{bmatrix}=\begin{bmatrix}
A & F\sqrt{1-e^2} \\
B & G\sqrt{1-e^2}
\end{bmatrix}.
\end{align}

Since for arbitrary values of $a,\Omega,i,\omega$, the components of $M'_{ij}$ can obtain any values, so can the four components of $M_{ij}$, and they can be treated as independent variables.

Equations \eqref{eq:xyM} and \eqref{eq:Kepler} approach a regular limit at $e\rightarrow 1$ far away from pericenter passage, implying a degeneracy when fitting such data. In particular, different values of $A,F,B,G$ and $e$ that result in the same values of $M_{ij}$ will given very similar motions on the sky as a function of time as long as $e$ is close to unity. 

Using equations \eqref{eq:MM'} and \eqref{eq:MpOmiom}, the relation between the eccentricity $e$ and the semi-major axis $a$ for fixed values of $M_{ij}$ can be expressed as
\begin{equation}\label{eq:eadegeneracy}
e=\sqrt{1-\frac{D^2-Ra^2}{a^2(L-a^2)}},
\end{equation}
where $D,L,R$ are constants given by
\begin{align}
&D=\det (\mathbf{M})=\sqrt{1-e^2}(AG-BF),\cr
&L=M_{xx}^2+M_{yx}^2=A^2+B^2,\cr
&R=M_{xy}^2+M_{yy}^2=(F^2+G^2)(1-e^2).
\end{align}

As we saw in the cases of $\zeta$ Boo and $\eta$ Oph in \ref{subsec:orbital_fit}, adding the VLTI/GRAVITY pericenter data broke the degeneracy. This happens because
close to the peri-center passage, the trajectory and the time dependence are sensitive to small changes in $e$. In particular, Kepler's Equation implies 
\begin{align}
\label{eq:dE_de}
\frac{dE}{de} = \frac{\sin(E)}{1-e \cos(E)}
\end{align}
\noindent so that for $e \approx 1$ the sensitivity of $E$ on $e$ at a given time is very high near pericenter and becomes weak away from pericenter. Additionally at the vicinity of the pericenter the sky positions are sensitive to relative changes in $1-e$. In particular, at the pericenter $E=0$ the separation on the sky is given by 
\begin{align}
&x=M_{xx}(1-e)\cr
&y=M_{yx}(1-e).
\end{align}

Nearly all of the highly eccentric visual binaries do not have constraining (and most often none) measurements near pericenter due to a combination of their long periods, small fraction of time spent near pericenter and the need for very high resolution techniques (which have become available only in the last decades) to resolve the binary near pericenter. As a result, only their four $M_{ij}$ can be well constrained, leading to strong degeneracies between the five orbital parameters that map to them (namely, $e$, $a$, $i$, $\omega$, $\Omega$). In practice, the lack of pericenter observations also leads to poor constraints on $T_p$, leading to further degeneracies between $e$, $T_p$ and possibly $P$ (depending on the orbital coverage) as well. The degeneracy is much less concerning for orbits that are not very eccentric because as can be seen from equation \ref{eq:dE_de} the eccentric anomaly then becomes most sensitive to $e$ at orbital phases other than pericenter (and in any case their pericenter passages are much less squeezed in time). Furthermore, the physical implications for a given error $\Delta e$ are much more important for a highly eccentric orbit than for a weakly eccentric one given the pericenter distance $a_p = a (1-e)$. 

\section{Discussion}
\label{sec:discussion}

\subsection{The degeneracy in the cases of $\zeta$ Boo and $\eta$ Oph}
\label{subsec:degeneracy_specific}

In Table \ref{table:orbital_fit_results} we report the $M_{ij}$ values for our orbital solutions for $\zeta$ Boo and $\eta$ Oph. As can be seen, their values are mostly consistent (within 2$\sigma$) between the solutions with and without including the VLTI/GRAVITY pericenter data, which is in accordance with our conclusion in Section \ref{sec:high_e_degeneracy} that $M_{ij}$ values should be quite robust to the lack of pericenter data. 

The predicted degeneracy lines based on equation \eqref{eq:eadegeneracy} are shown in red lines in Figures \ref{fig:a_e_degeneracy_zeta_Boo} and \ref{fig:a_e_degeneracy_eta_oph} using the best fit $M_{ij}$ values for each case. As can be seen, for both targets the distributions for the fit without the pericenter data follow the predicted degeneracy lines, while in the fit including the pericenter data the degeneracy is broken.

However, there is an important difference between the cases of $\zeta$ Boo and $\eta$ Oph. While in the case of the latter the degeneracy only led to higher uncertainties and stronger correlations for the solution without the pericenter data, in the case of $\zeta$ Boo the degeneracy also led to a bias which, in the absence of the pericenter data and without a clear understanding of the degeneracy, would have led to wrong conclusions on the orbital parameters. 

In order to further understand the case of $\zeta$ Boo, we experimented with including different combinations of the VLTI/GRAVITY epochs in the fit and found that the post-pericenter (fourth) epoch is by far the most important one to drive the fit towards the correct solution. Figure \ref{fig:a_vs_e_zeta_boo_3pochs} in Appendix \ref{app:zeta_boo_additional} shows the resulting $a$-$e$ distribution when only including the first three (pre-pericenter) VLTI/GRAVITY epochs. It is clear that the degeneracy is still present and that the solution is still biased towards high eccentricities. Figure \ref{fig:gravity_fits_zeta_boo_prediction} shows the predicted interferometric data (black lines) for the fourth VLTI/GRAVITY epoch based on this solution, which is clearly inconsistent with the actual data (colored) -- the predicted position $(\Delta \alpha_*, \Delta \delta) = (-34.9, 105.9) \text{ mas}$ is offset by 26.4 mas compared to the measured one in Table \ref{table:gravity_results_zeta_boo}. From this we conclude that having an epoch post-pericenter is crucial to converge to the correct solution. 

\begin{figure}[]
x \includegraphics[width=\columnwidth]{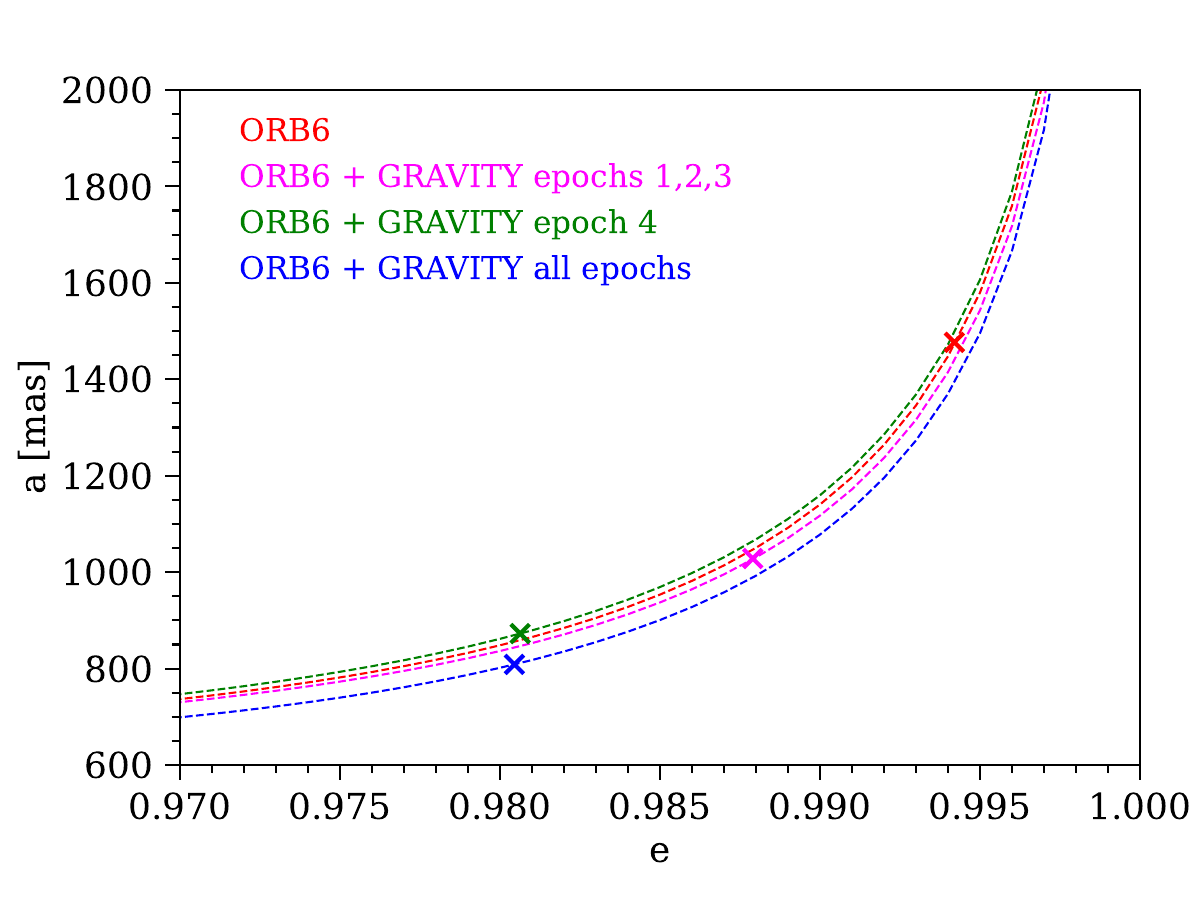}
 \caption{\label{fig:fit_orbit_zeta_boo_a_vs_e_many_fits} Best-fit (a,e) solutions for $\zeta$ Boo using different data combinations. The dashed lines show the predicted degeneracy lines.}
\end{figure}

Figure \ref{fig:fit_orbit_zeta_boo_a_vs_e_many_fits} summarizes our findings by showing the best fit values for $a$ and $e$ for different data combinations, namely using only the ORB6 data, only the first three epochs (pre-pericenter) of VLTI/GRAVITY, only the fourth (post-pericenter) epoch and all the epochs. The dashed lines show the predicted degeneracy line using equation \ref{eq:eadegeneracy} based on the best-fit $M_{ij}$ values for each case. It is clear that the solutions fall along the degeneracy lines and that the post-pericenter VLTI/GRAVITY epoch has the most important contribution to breaking the degeneracy. 

Our explanation for the bias in the $\zeta$ Boo solution without the pericenter data is that unavoidable imperfections in the assumptions (such as systematic errors and unaccounted data correlations) can make the fit extremely sensitive to the presence of strong degeneracies and correlations in the case of very eccentric orbits that lack a decent pericenter coverage.

\subsection{Wide companions and future evolution}
\label{subsec:wide_companion}

In \cite{Waisberg24} we suggested that the wide 41.3 kau companion to $\zeta$ Boo could have been responsible for its very high eccentricity. Even though $\zeta$ Boo is not experiencing eccentricity oscillations at the present time (the outer companion is too far away), we proposed a scenario in which the outer companion could have been much closer in the past but has since migrated outwards on successively more eccentric orbits by stealing energy from the inner binary \citep[e.g. ][]{Mushkin20}. At large enough outer semi-major axis the Galactic tidal field could eventually decouple the system and freeze the high eccentricity of $\zeta$ Boo \citep[see ][for details]{Waisberg24}. 

For $\eta$ Oph the individual proper motions measured by Hipparcos and the magnitude difference $\Delta H_p = 0.512$ imply a proper motion $(v_{\alpha*},v_{\delta}) = (36.9\pm1.1,94.0\pm0.7) \text{ mas}\text{ yr}^{-1}$ for the photocenter at the Hipparcos epoch (1991.25). Correcting for the orbital motion of the photocenter at this epoch based on the orbital parameters measured in this paper, $\Delta H_p$ and the mass ratio $q\approx0.90$, we estimate the system (center of mass) linear proper motion to be $(v_{\alpha*},v_{\delta}) = (37.6\pm1.1,95.1\pm0.7) \text{ mas}\text{ yr}^{-1}$. As we did for $\zeta$ Boo, we searched for companions to $\eta$ Oph that would share its parallax and proper motion in the Gaia DR3 catalog up to a projected separation of 5 deg and did not find any candidate. The lack of a third companion to $\eta$ Oph could mean that its high eccentricity is natal or that it is a result of a three-body interaction in which one of the stars was ejected or still that an outer companion excited Kozai-Lidov oscillations in the past but has since migrated outwards and became unbound.

As the future evolution of both $\zeta$ Boo and $\eta$ Oph is unlikely to be affected by additional stars, they will likely circularize during the giant or AGB phase of the primary while conserving orbital angular momentum (the spin angular momenta are negligible compared to the orbital) to a semi-major axis of $a_f = a_i (1-e_i^2)$, which is $a_f = 4.2 \text{ au}$ for $\eta$ Oph and $a_f = 1.7 \text{ au}$ for $\zeta$ Boo. Follow-up evolution will be similar to other binaries at similar separations and masses.

\subsection{Which is the most eccentric binary known?}
Our measured eccentricity $e=0.980450\pm0.000064$ for $\zeta$ Boo confirms it as a possible candidate for the most eccentric binary known. The current record holder among spectroscopic binaries appears to still be the G9V+M0V binary (P=890 d, a=2.1 au) Gliese 586A (HIP 75718) with $e=0.97608\pm0.00004$ \citep{Tokovinin91,Duquennoy92,Strassmeier13}, followed closely by the F7V+F7V binary (P=1247 d, a=3.3 au) 41 Dra (HIP 88136) with $e=0.9754\pm0.0001$ \citep{Tokovinin03}. Gaia DR3 reported only one spectroscopic binary with $e>0.97$, namely UCAC4 120-012867 (Gaia DR3 source 5287239332770307072) with $e=0.973\pm0.009$ and period P=87.86 d. There are a further nine stars in Gaia DR3 with astrometric orbital solutions with $e>0.97$ and periods between 300 and 6318 d, but most of them have relatively large errors ($\sigma_e = 0.05-0.2$) with the exception of CD-54 7681 (Gaia DR3 6653167403963974016) with $e=0.972\pm0.013$ and $P=4598\pm748$ d and Gaia DR3 1792404030031405952 with $e=0.973\pm0.014$ and $P=6318\pm1702$ d, both of which probably require validation given their large periods compared to the span of the Gaia data. 

Within ORB6\footnote{Version updated on Nov 20 05:58:11 EST 2024.} there are 37 reported binaries with $0.97 \leq e < 1$. After excluding seven systems with ORB6 grades greater than 5.0, one system that was solved by Gaia and found to have a much lower eccentricity, the two aforementioned spectroscopic binaries (which also have visual orbits) and two S-stars in the Galactic center, we were left with 25 extremely eccentric binary candidates. Table \ref{table:high_e_ORB6} lists the properties of such systems and more details about the excluded systems can be found in Appendix \ref{app:high_e_orb6}. A closer look at these candidates reveals many issues:
\begin{enumerate}
\item 15 of the solutions do not report uncertainties to the orbital parameters; 
\item none of the solutions have an ORB6 grade smaller than 2.0 and only 8 of them have a grade smaller than 3.0\footnote{The ORB6 grades are assigned based on the orbital phase coverage and the quality of the orbital solution (the smaller the grade the better). Their definitions can be found \href{https://www.astro.gsu.edu/wds/orb6/orb6text.html\#grading}{in this link}.};
\item in the majority of the solutions, the resulting dynamical mass is too high (in most cases by an order of magnitude or more) compared to the expected total mass based on the spectral type of the primary and the magnitude difference; 
\item for 9 of the systems there are alternative solutions in the master file of ORB6 with similar grade, lower $e$ and lower $a$ (resulting in lower and often more reasonable dynamical masses). 
\end{enumerate}

Such a high fraction of problematic solutions are not surprising in light of the degeneracy intrinsic to highly eccentric orbits discussed in Section \ref{sec:high_e_degeneracy} and considering that nearly all of the solutions suffer from limited orbital coverage and a lack of observations near periastron passage. A detailed analysis of each solution is beyond the scope of this paper, but from visually looking at the orbital plots available for each system in ORB6 and based on our experience we found that there is only one case for which the very high eccentricity is very likely robust, namely WDS 05354-3316 (HIP 26245) with an F6V primary, $P=130\pm4$ yr and $e=0.985\pm0.002$, for which speckle interferometry observations with reported precision below 1 mas were obtained for about a decade close to pericenter passage \citep{Tokovinin20} and for which the resulting dynamical mass is reasonable. There are three other cases with reasonable dynamical masses that do not have observations close to pericenter nor reported uncertainties (namely WDS 13038-2035 with $e=0.988$, WDS 01077-1557 with $e=0.98$ and WDS 14411+1344 = Zeta Bo\"otis with $e=0.98$). 

To the best of our knowledge, $\zeta$ Boo has thus the second highest well constrained eccentricity of all known binaries, with the highest belonging to WDS 05354-3316 (HIP 26245). It is likely that binaries with larger eccentricity lurk within the ORB6 data, but require additional observations to break the degeneracy. 

Finally, we note that besides high angular resolution observations throughout the pericenter passage, there are a few other alternatives to break the degeneracy and obtain an accurate value for $e$ in very eccentric visual binaries. One is to obtain radial velocity measurements during the pericenter passage, although in practice this may be challenging for stars of intermediate mass and above which frequently have broad lines due to large rotational velocities. Another approach is to use dynamical mass constraints in order to constrain $a$ and therefore obtain $e$, $i$, $\omega$ and $\Omega$ from the four $M_{ij}$. However, in addition to not being an ideal approach (as one would aspire to have orbital parameters that do not depend on external constraints on the component masses), this requires an accurate and precise parallax (which is often an issue for close visual binaries) and might not be very useful for more massive stars for which photometric or spectroscopic masses are much less reliable. It is clear, however, that external dynamical mass constraints can be quite useful for identifying problematic solutions (as we have done for $\zeta$ Boo in this paper). 

\section*{Acknowledgments}

We thank the anonymous referee for their comments and suggestions. This research has made use of the Washington Double Star Catalog maintained at the U.S. Naval Observatory. In particular, we thank Dr. Rachel Matson for providing us with the individual ORB6 astrometric data files of $\zeta$ Boo and $\eta$ Oph.

This work has made use of data from the European Space Agency (ESA) mission {\it Gaia} (\url{https://www.cosmos.esa.int/gaia}), processed by the {\it Gaia} Data Processing and Analysis Consortium (DPAC,
\url{https://www.cosmos.esa.int/web/gaia/dpac/consortium}). Funding for the DPAC
has been provided by national institutions, in particular the institutions
participating in the {\it Gaia} Multilateral Agreement. This research has made use of the Jean-Marie Mariotti Center \texttt{SearchCal} service co-developped by LAGRANGE and IPAG, the CDS Astronomical Databases SIMBAD and VIZIER, NASA's Astrophysics Data System Bibliographic Services, NumPy \citep{van2011numpy} and matplotlib, a Python library for publication quality graphics \citep{Hunter2007}.

\section*{Data availability}
The VLTI/GRAVITY data underlying this article are publicly available from the ESO archive. Reduced data can be provided by the corresponding author on request.

\bibliographystyle{aasjournal}
\bibliography{main}{}

\appendix

\section{A. Additional interferometric fits}
\label{app:additional_fits}

\begin{figure*}[]
\centering
\includegraphics[width=\textwidth]{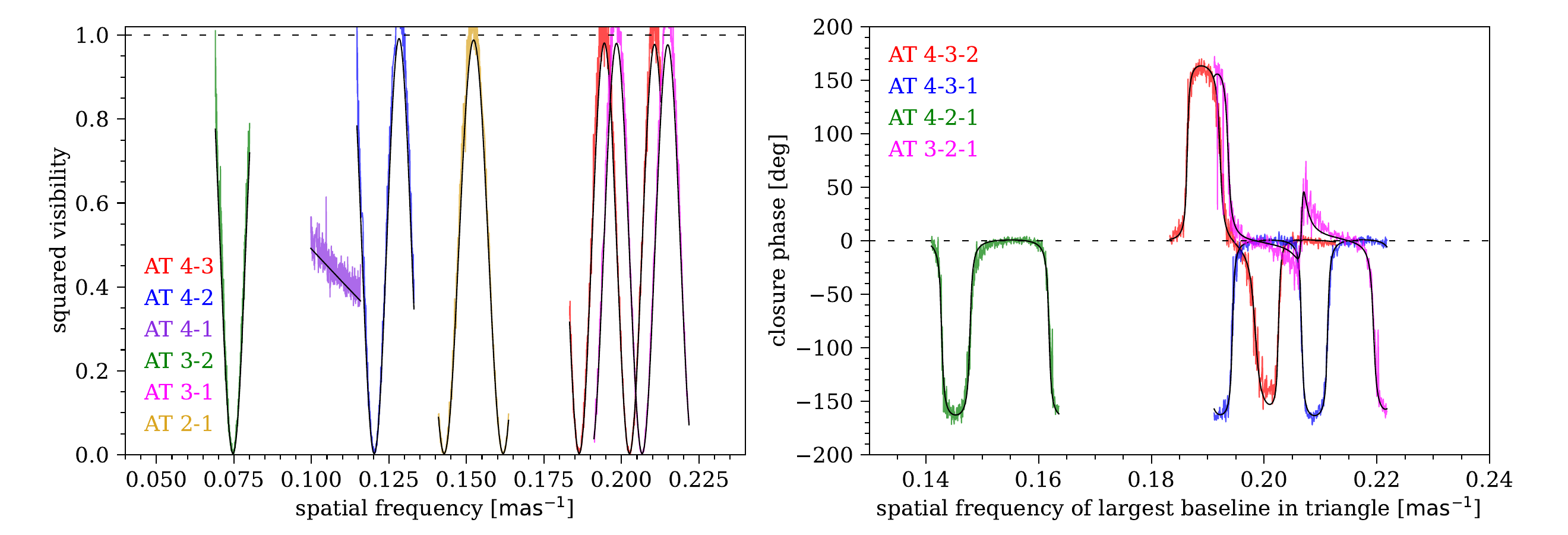} \\
\includegraphics[width=\textwidth]{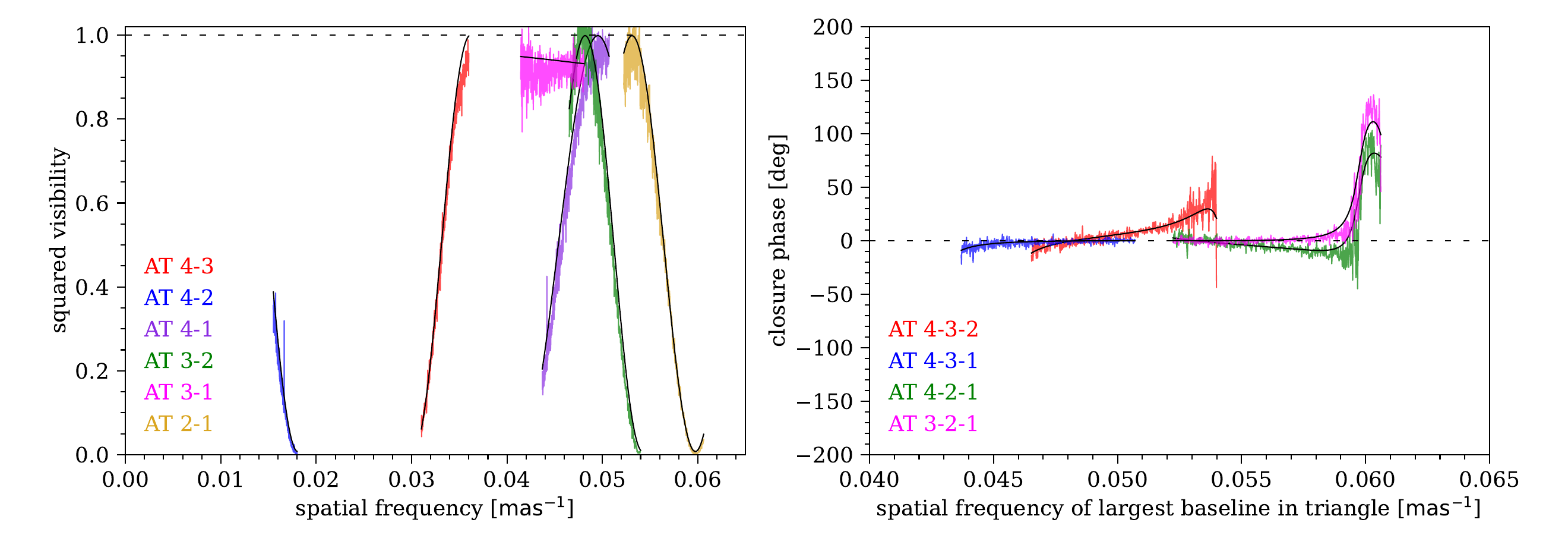}
\caption{\label{fig:gravity_fits_zeta_boo_additional} VLTI/GRAVITY data (colored) and best fit binary model (solid black) for Epochs 2 (top) and 4 (bottom) of Zeta Bo\"otis.}
\end{figure*}

\begin{figure*}[]
\centering
\includegraphics[width=\textwidth]{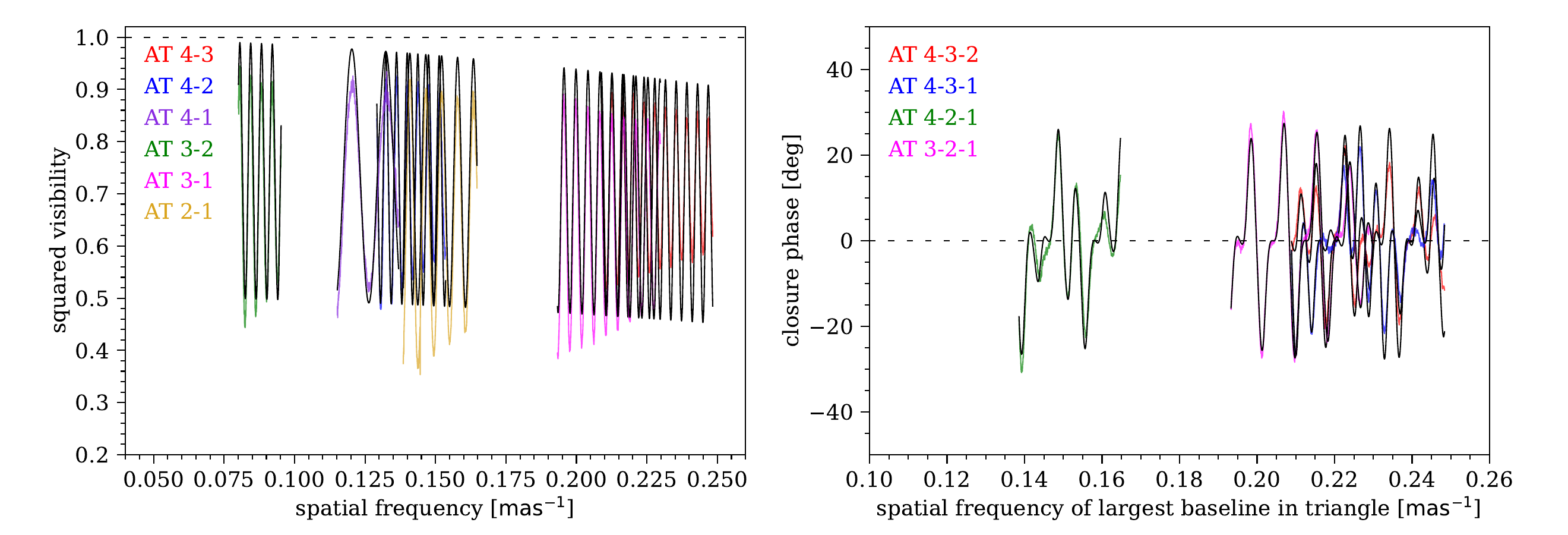}\\
\includegraphics[width=\textwidth]{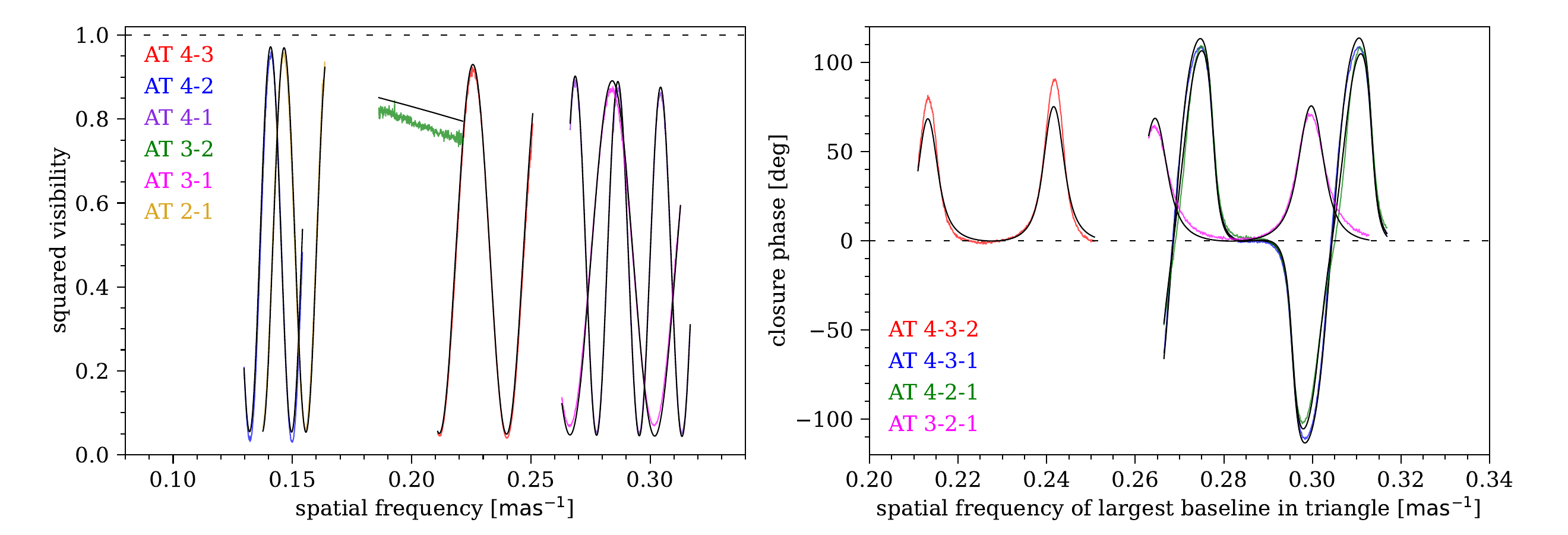}\\
\includegraphics[width=\textwidth]{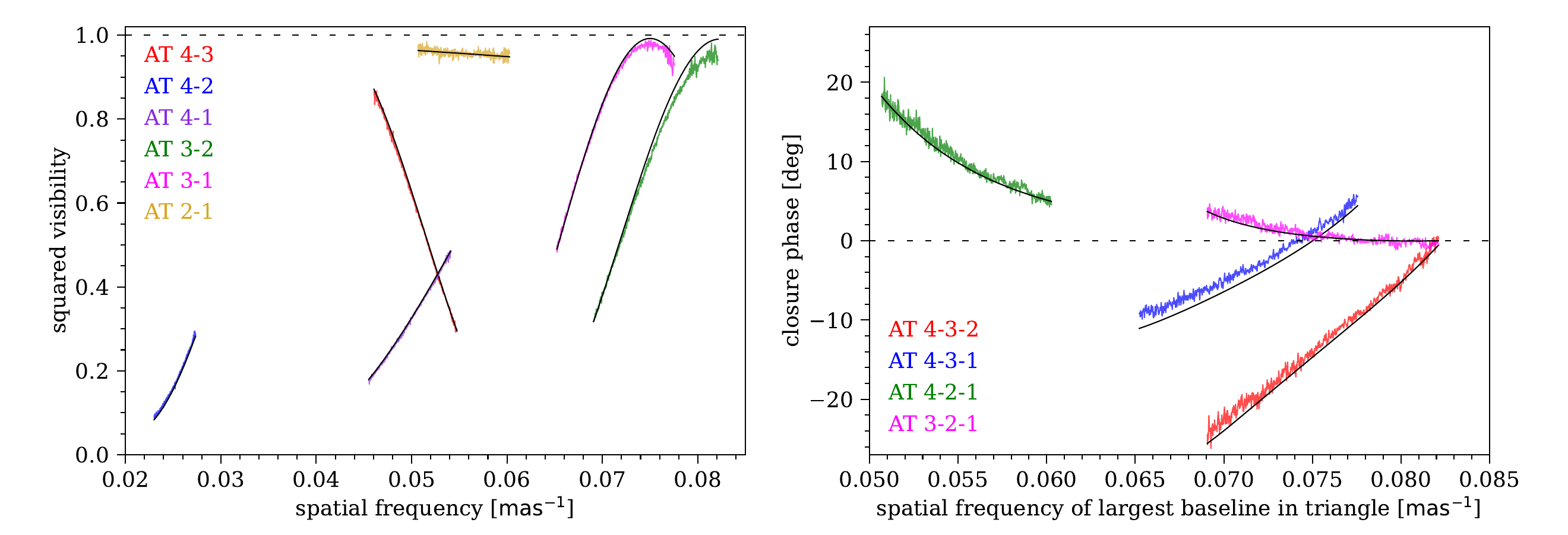}\\
\includegraphics[width=\textwidth]{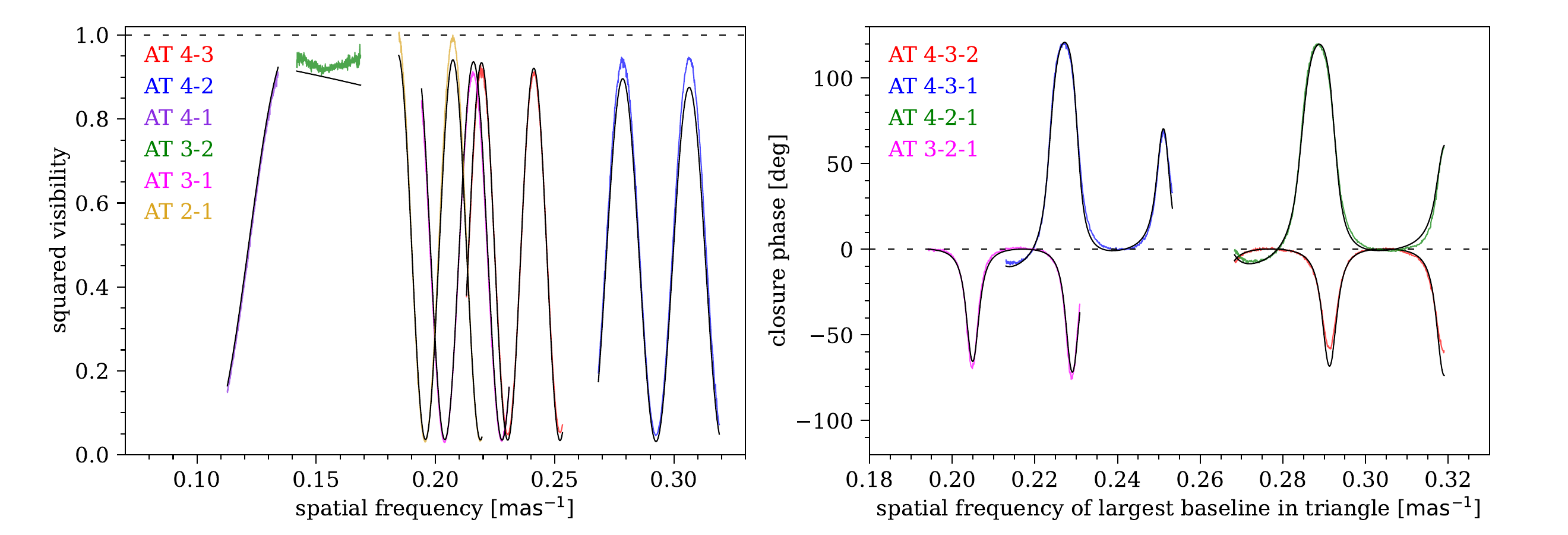}
\caption{\label{fig:gravity_fits_eta_oph_additional} VLTI/GRAVITY data (colored) and best fit binary model (solid black) for Epochs 1, 2, 5 and 6 (top to bottom) of Eta Ophiuchi.}
\end{figure*}

Figures \ref{fig:gravity_fits_zeta_boo_additional} and  \ref{fig:gravity_fits_eta_oph_additional} show the VLTI/GRAVITY interferometric data and best fit binary models for the remaining observation epochs. 

\section{B. Data selection and error estimation for ORB6 data}
\label{app:orb6_details}

Here we summarize our data selection and error estimation when using the ORB6 astrometric data. We started by excluding a few old measurements which did not have both a position angle (PA) and a separation and flipping some PA that were clearly offset by $180^{\degr}$. A lot of the measurements (especially the older ones) do not have reported errors; furthermore, many of the reported errors are underestimated since as a rule they do not take into account systematic errors. In order to have a better estimate of the errors to appropriately combine the data, we separated the ORB6 data into a group of low-resolution observations (the vast majority of which are micrometry observations before 1980) and another of high-resolution observations (which include some visual interferometry before 1980 and the vast majority of which are speckle interferometry, aperture masking and adaptive optics after 1980). For $\zeta$ Boo, we also created a separate group for the interferometric observations with the Palomar Testbed Interferometer \citep[PTI; ][]{Muterspaugh10a}, which are also included in the ORB6 data. For each group, we then found the best-fit orbit (with all the epochs having equal weight) and scaled the errors so that the average residual squared is equal to the square of the error\footnote{We note that we also attempted to divide the high-resolution observations into two subgroups -- namely, observations before and after 2004 -- in order to take into account the improvement of the instruments and of the astrometric precision over some decades. However, we found that the resulting errors were very similar between the two subgroups, suggesting that at least for our two targets the errors are ultimately dominated by systematics.}. Within each group, we then excluded blatant outliers (defined as the residual in either $\Delta \alpha_*$ or $\Delta \delta$ being larger than four times the estimated error). Finally, the orbit fitting was repeated and the final error was estimated as described above.

\section{C. Additional plots for $\zeta$ Boo}
\label{app:zeta_boo_additional}

\begin{figure*}[]
\centering
\includegraphics[width=0.5\textwidth]{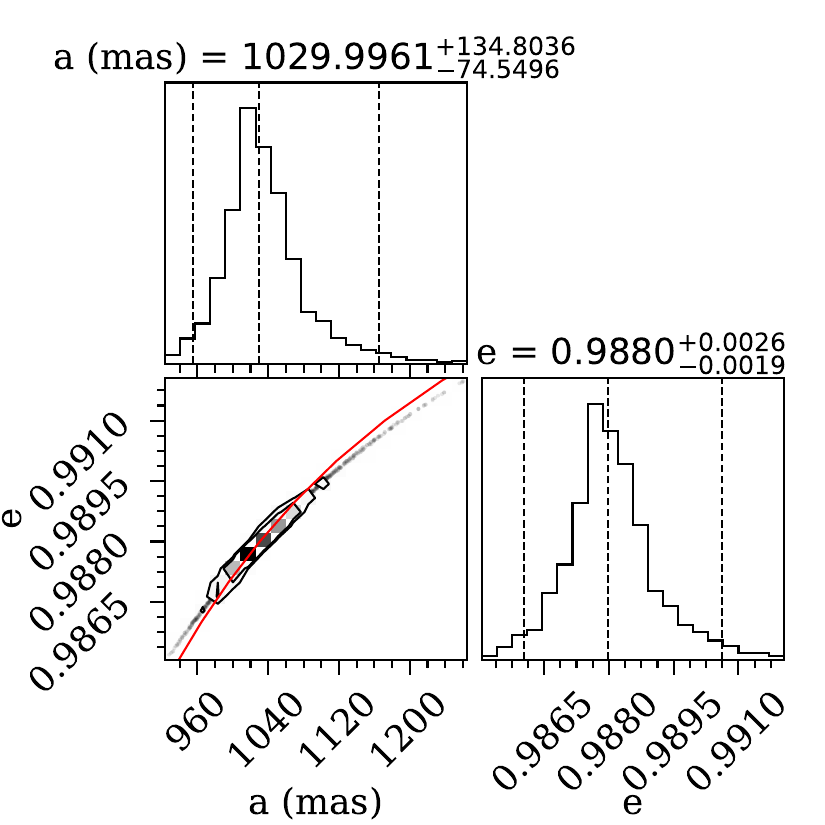}
\caption{\label{fig:a_vs_e_zeta_boo_3pochs} $a$-$e$ distribution for the orbital solution combining the ORB6 data with the first three (pre-pericenter) VLTI/GRAVITY epochs. The red line shows the predicted degeneracy based on equation \ref{eq:eadegeneracy}.}
\end{figure*}

\begin{figure*}[]
\centering
\includegraphics[width=\textwidth]{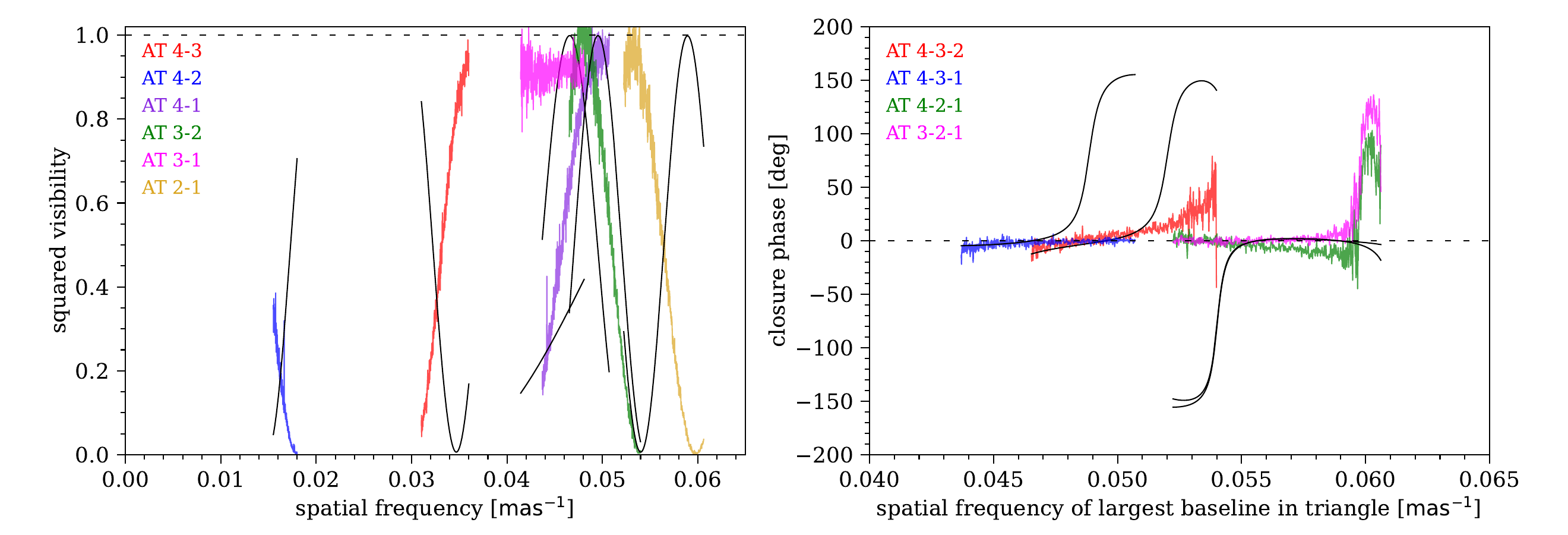}
\caption{\label{fig:gravity_fits_zeta_boo_prediction} Predicted interferometric data (solid black) for the fourth (post-pericenter) VLTI/GRAVITY epoch based on the orbital solution combining the ORB6 data with the first three (pre-pericenter) VLTI/GRAVITY epochs compared to the measured data (colored). The predicted position $(\Delta \alpha_*, \Delta \delta) = (-34.9, 105.9) \text{ mas}$ is off by 26.4 mas compared to the measured position in Table \ref{table:gravity_results_zeta_boo}.}
\end{figure*}

Figures \ref{fig:a_vs_e_zeta_boo_3pochs} and \ref{fig:gravity_fits_zeta_boo_prediction} show additional plots that help to understand the inconsistency between the orbital fits of $\zeta$ Boo as detailed in \ref{subsec:degeneracy_specific}.

\section{D. Extremely eccentric binary candidates from ORB6}
\label{app:high_e_orb6}

In Table \ref{table:high_e_ORB6} we list the orbital solutions in ORB6 (version updated on Nov 20 05:58:11 EST 2024) which have $e \geq 0.97$. The distances are based on the inverse parallax from Gaia DR3, Gaia DR2 or Hipparcos (in this order of preference). The spectral types are taken from SIMBAD. We have excluded from this list: 

\begin{enumerate}
\item Seven systems (WDS 01522-5220, WDS 15155+3319, WDS 14330+0656, WDS 13344-4224, WDS 02231+7021, WDS 07192+2058 and WDS 16202-3734) that have ORB6 grades greater than 5.0, as in such cases the orbital parameters are essentially unconstrained and a large eccentricity may even reflect that the companion is an unrelated background or foreground object. 

\item WDS 22204+4625 (HIP 110291), which is a Hipparcos astrometric binary with a solution $e=0.99\pm0.09$ ($P=794\pm52$ d) but whose SB1 solution in Gaia DR3 revealed a much lower $e=0.48\pm0.05$ ($P=736\pm14$ d).

\item WDS 18002+8000 (41 Dra) and WDS 15282-0921 (Gliese 586A). These are spectroscopic binaries and their eccentricities are very well constrained from their radial velocity curves during periastron passage. They have also been resolved through high angular resolution observations. It is interesting to note that their orbital solutions have ORB6 grades of 2.3 and 2.0 respectively. 

\item WDS 17457-2900 SgrA* - S175 ($P=96.2\pm5.0$ \text{ yr}; $e=0.9867\pm0.0018$) and SgrA* - S14 ($P=55.3\pm0.5$ \text{ yr}; $e=0.9761\pm0.0037$). These are stars orbiting the supermassive black hole SgrA* at the Galactic center. Their orbits have been intensely monitored with Adaptive Optics for about three decades \citep{Gillessen17}. The very high eccentricities are very likely real but they are not stellar binaries. 
\end{enumerate}

\begin{table*}
\centering
\caption{\label{table:high_e_ORB6} Very high eccentricity solutions ($e \geq 0.97)$ with ORB6 grades smaller or equal to 5.0.}
\begin{tabular}{ccccccccc}
\hline \hline
\shortstack{WDS\\other} & \shortstack{distance\\(pc)} & e & \shortstack{Period\\(yr)} & \shortstack{a(")\\(au)} & \shortstack{spectral type\\$\Delta m$} & \shortstack{$M_{\mathrm{dyn}}$\\($M_{\odot}$)} & $T_p$ & ORB6 grade \\ [0.3cm]

\shortstack{18531-5011\\HIP 92680\\PZ Tel}\tablenotemark{a} & $47.25\pm0.05$ & 0.996 & $1766.6$ & \shortstack{3.02\\142.7} & \shortstack{G9IV\\$\Delta K = 4.0$}  & 0.93 & 2001.7 & 4.3 \\ [0.3cm]

\shortstack{15420+4203\\HIP 76882} & $86\pm5$ & 0.991 & $70.59$ & \shortstack{0.263\\22.6} & \shortstack{F8\\$\Delta V = 0.6$}  & 2.4 & 1993.39 & 4.0 \\ [0.3cm]

\shortstack{00508+3203\\BD+31 127} & $123\pm3$ & 0.99 & $106.44$ & \shortstack{0.786\\96.7} & \shortstack{F8\\$\Delta V = 0.16$}  & 80 & 2058.04 & 4.4 \\ [0.3cm]

\shortstack{10281+4847\\HIP 51248} & $22.9\pm0.2$ & $0.99\pm0.02$ & $80.59\pm3.44$ & \shortstack{$8.46\pm3.32$\\$193.7\pm76.0$} & \shortstack{F9V\\$\Delta V = 6.2$}  & $1121\pm1628$ & $ 1981.11\pm5.17$ & 4.7 \\ [0.3cm]

\shortstack{22248+2233\\HIP 110640} & $20.87\pm0.01$ & $0.990\pm0.008$ & $129.24\pm8.32$ & \shortstack{$2.48\pm0.61$\\$193.7\pm76.0$} & \shortstack{K7\\$\Delta V = 2.53$}  & $8.6\pm7.2$ & $ 1973.19\pm2.15$ & 4.3 \\ [0.3cm]

\shortstack{23409+2022\\HIP 116838} & $25.49\pm0.04$ & $0.99\pm0.21$ & $124.90\pm5.11$ & \shortstack{$4.34\pm3.35$\\$110.6\pm85.4$} & \shortstack{K3.5V\\$\Delta V = 2.31$}  & $87\pm411$ & $ 1963.40\pm7.43$ & 4.6 \\ [0.3cm]

\shortstack{13117-2633\\HIP 64375}\tablenotemark{a} & $110.6\pm6.9$ & $0.989\pm0.008$ & $20.07\pm0.50$ & \shortstack{$0.226\pm0.002$\\$25.0\pm0.2$} & \shortstack{A5V\\$\Delta V = 0.23$}  & $38.8\pm7.9$ & $ 2021.80\pm0.80$ & 2.2 \\ [0.3cm]

\shortstack{13038-2035\\HIP 63738} & $28.2\pm0.7$ & $0.988$ & $58.98$ & \shortstack{$0.718$\\$20.2$} & \shortstack{F8.5V\\$\Delta V = 0.26$}  & $2.4$ & $ 1964.10$ & 2.7 \\ [0.3cm]

\shortstack{21426+4103\\HIP 107177}\tablenotemark{a} & $120.5\pm10.3$ & $0.986$ & $106.9$ & \shortstack{$0.673$\\$81.1$} & \shortstack{F2V\\$\Delta V = 0.52$}  & $46.7$ & $ 1947.53$ & 2.6 \\ [0.3cm]

\shortstack{05354-3316\\HIP 25245} & $47.2\pm0.6$ & $0.985\pm0.002$ & $130.0\pm4.0$ & \shortstack{$0.659\pm0.052$\\$31.1\pm2.5$} & \shortstack{F6V\\$\Delta V = 1.69$}  & $1.8\pm0.4$ & $2019.04\pm0.03$ & 3.1 \\ [0.3cm]

\shortstack{10120-0612\\HIP 49961} & $85.1\pm10.5$ & $0.984$ & $256.04$ & \shortstack{$0.433$\\$36.9$} & \shortstack{F5V\\$\Delta V = 0.40$}  & $0.76$ & $1942.68$ & 4.2 \\ [0.3cm]

\shortstack{10269+1713\\HIP 51145}\tablenotemark{a} & $86.9\pm1.7$ & $0.984$ & $139.8$ & \shortstack{$0.470$\\$40.8$} & \shortstack{F8\\$\Delta V = 0.73$}  & $3.5$ & $1966.98$ & 2.7 \\ [0.3cm]

\shortstack{10131+2725\\HIP 50052}\tablenotemark{a} & $67.9\pm0.1$ & $0.981\pm0.009$ & $192.9\pm12.4$ & \shortstack{$1.27\pm0.42$\\$86.2\pm28.5$} & \shortstack{G5V\\$\Delta V = 1.21$}  & $17\pm20$ & $1941.3\pm2.2$ & 3.3 \\ [0.3cm]

\shortstack{01077-1557\\HIP 5295} & $110.1\pm2.2$ & $0.98$ & $15.51\pm1.00$ & \shortstack{$0.074\pm0.020$\\$8.1\pm2.2$} & \shortstack{G1V\\$\Delta V = 0.63$}  & $2.2\pm2.2$ & $2019.12\pm0.25$ & 2.8 \\ [0.3cm]

\shortstack{04170+1941 \\HIP 19975} & $80.6\pm0.8$ & $0.98$ & $63.3$ & \shortstack{$0.358$\\$28.9$} & \shortstack{F5\\$\Delta V = 1.68$}  & $6.0$ & $1972.4$ & 2.9 \\ [0.3cm]

\shortstack{09357+3549\\HIP 47080}\tablenotemark{a} & $11.24\pm0.01$ & $0.98\pm0.01$ & $241.1\pm64.0$ & \shortstack{$7.18\pm2.64$\\$80.4\pm29.6$} & \shortstack{G8Va\\$\Delta V = 7.7$}  & $9\pm30$ & $1941.7\pm1.3$ & 5.0 \\ [0.3cm]

\shortstack{14094+1015\\HIP 69160} & $59.4\pm1.2$ & $0.98$ & $8.36$ & \shortstack{$0.104$\\$6.2$} & \shortstack{G3V\\$\Delta V = 0.98$}  & $3.4$ & $2018.88$ & 3.4 \\ [0.3cm]

\shortstack{14411+1344\\HIP 71795\\Zeta Bo\"otis} & $51.7\pm0.2$ & $0.98$ & $125.24$ & \shortstack{$0.825$\\$42.7$} & \shortstack{A1V\\$\Delta V = 0.09$}  & $4.93$ & $1898.58$ & 2.2 \\ [0.3cm]

\shortstack{18575+5814\\HIP 93068}\tablenotemark{a} & $123.3\pm0.9$ & $0.98$ & $261.6$ & \shortstack{$1.051$\\$129.6$} & \shortstack{A2V\\$\Delta V = 0.43$}  & $31.8$ & $1885.89$ & 3.0 \\ [0.3cm]

\shortstack{23241+5732\\HIP 115529} & $300\pm54$ & $0.98$ & $300$ & \shortstack{$0.296$\\$22$} & \shortstack{B2V\\$\Delta V = 0.27$}  & $7.7$ & $1935.2$ & 3.5 \\ [0.3cm]

\shortstack{18534+2553\\HD 337117} & $143.3\pm16.2$ & $0.978$ & $175$ & \shortstack{$0.74$\\$63.6$} & \shortstack{F5\\$\Delta V = 1.0$}  & $38.9$ & $1968.3$ & 4.0 \\ [0.3cm]

\shortstack{12409+2708\\HIP 61890}\tablenotemark{a} & $117.0\pm8.9$ & $0.976$ & $38.13$ & \shortstack{$0.118$\\$8.3$} & \shortstack{F5V\\$\Delta V = 0.2$}  & $1.8$ & $1988.88$ & 3.2 \\ [0.3cm]

\shortstack{03151+1618\\HIP 15134} & $73.6\pm2.9$ & $0.975\pm0.006$ & $100.5\pm4.6$ & \shortstack{$0.442\pm0.045$\\$19.5\pm2.0$} & \shortstack{G0\\$\Delta V = 0.5$}  & $3.4\pm1.2$ & $1991.30\pm0.69$ & 3.9 \\ [0.3cm]

\shortstack{17161+2316\\HIP 84468} & $263\pm17$ & $0.975\pm0.004$ & $90.50\pm1.00$ & \shortstack{$0.506\pm0.0055$\\$79.9\pm0.8$} & \shortstack{RGB\\$\Delta V = 0.8$}  & $288\pm61$ & $2012.83\pm0.05$ & 3.9 \\ [0.3cm]

\shortstack{08507+1800\\HIP 43421}\tablenotemark{a} & $361\pm134$ & $0.972\pm0.010$ & $113.4\pm1.0$ & \shortstack{$0.979\pm0.040$\\$212$} & \shortstack{G5\\$\Delta V = 0.09$}  & $3435\pm7e7$ & $2018.01\pm0.50$ & 2.5 \\ 

\hline \\ 
\end{tabular}
\tablenotetext{1}{There are alternative solutions in the master file of ORB6 with lower $e$.}

\end{table*}

For several of these systems there are additional solutions in the master file of ORB6 with lower eccentricities and often more reasonable dynamical masses: 

WDS 18531-5011: There is an alternative solution in the master file of ORB6 with a grade 4.2 and lower $e=0.965$.

WDS 13117-2633: There are several alternative solutions in the master file of ORB6 with grades between 2.1 and 3.2, lower $e \sim 0.93$ (and even $e\sim 0$ with half the orbital period) and more reasonable dynamical masses.

WDS 21426+4103: There is alternative solution in the master file of ORB6 with grade 3.0, lower $e=0.979$ and a smaller but still problematic dynamical mass $M_{\mathrm{dyn}}=8.7 M_{\odot}$.

WDS 10269+1713: There is an alternative solution in the master file of ORB6 with grade 3.0, $e=0.965$ and a more reasonable dynamical mass $M_{\mathrm{dyn}}=2.6 M_{\odot}$.

WDS 10131+2725: There is an alternative solution in the master file of ORB6 with grade 3.5, $e=0.958$ and a smaller but still problematic dynamical mass $M_{\mathrm{dyn}}=5.8 M_{\odot}$.

WDS 09357+3549: There is an alternative solution in the master file of ORB6 with grade 5.1, $e=0.88$ and a more reasonable albeit still somewhat too high dynamical mass $M_{\mathrm{dyn}}=2.0 M_{\odot}$.

WDS 18575+5814: There is an alternative solution in the master file of ORB6 with grade 3.0, $e=0.900$ and a much more reasonable dynamical mass $M_{\mathrm{dyn}}=3.8 M_{\odot}$.

WDS 12409+2708: There is an alternative solution in the master file of ORB6 with grade 3.2 and the same dynamical mass but lower $e=0.956\pm0.010$.

WDS 08507+1800: There is an alternative solution in the master file of ORB6 with grade 3.2, much lower $e=0.139$ and a more reasonable dynamical mass $M_{\mathrm{dyn}}=24 M_{\odot}$ (the parallax is very uncertain).

\end{document}